\documentclass[conference]{IEEEtran}

\usepackage[colorlinks,urlcolor=blue,linkcolor=blue,citecolor=blue]{hyperref}
\usepackage{amsmath}
\usepackage{color,array}
\usepackage{subfig}
\usepackage[ruled,vlined]{algorithm2e}
\usepackage{graphicx}
\usepackage{multirow}

\ifCLASSINFOpdf
\else
\fi
\begin{document}
%

\title{DetectX- Adversarial Input \underline{Detect}ion using Current Signatures in Memristive \underline{X}Bar Arrays} 



\author{\IEEEauthorblockN{Abhishek Moitra}
\IEEEauthorblockA{Electrical Engineering department\\
Yale University\\
New Haven, CT, USA\\
Email: abhishek.moitra@yale.edu}
\and
\IEEEauthorblockN{Priyadarshini Panda}
\IEEEauthorblockA{Electrical Engineering department\\
Yale University\\
New Haven, CT, USA\\
Email: priya.panda@yale.edu}}


%


\maketitle

\begin{abstract}
Adversarial input detection has emerged as a prominent technique to harden Deep Neural Networks(DNNs) against adversarial attacks. Most prior works use neural network-based detectors or complex statistical analysis for adversarial detection. These approaches are computationally intensive and vulnerable to adversarial attacks. To this end, we propose DetectX - a hardware friendly adversarial detection mechanism using hardware signatures like Sum of column Currents (SoI) in memristive crossbars (XBar). We show that adversarial inputs have higher SoI compared to clean inputs. However, the difference is too small for reliable adversarial detection. Hence, we propose a dual-phase training methodology: Phase1 training is geared towards increasing the separation between clean and adversarial SoIs; Phase2 training improves the overall robustness against different strengths of adversarial attacks. For hardware-based adversarial detection, we implement the DetectX module using 32nm CMOS circuits and integrate it with a Neurosim-like analog crossbar architecture. We perform hardware evaluation of the Neurosim+DetectX system on the Neurosim platform using datasets-CIFAR10(VGG8), CIFAR100(VGG16) and TinyImagenet(ResNet18). Our experiments show that DetectX is 10x-25x more energy efficient and immune to dynamic adversarial attacks compared to previous state-of-the-art works. Moreover, we achieve high detection performance (ROC-AUC$>$0.95) for strong white-box and black-box attacks. The code has been released at \url{https://github.com/Intelligent-Computing-Lab-Yale/DetectX}
\end{abstract}


%
\IEEEpeerreviewmaketitle

\section{Introduction}
\label{introduction}
\IEEEPARstart{D}{eep} Neural Networks (DNNs) have been shown to be proficient in carrying out tasks like autonomous driving, image segmentation and real-time object detection. However, they have been shown to be vulnerable to adversarial attacks. Here, extremely small noise is added to an image to fool the DNN \cite{huang2017adversarial,dong2018boosting}. This prevents DNNs from being autonomously deployed in critical applications like medical diagnostics, aviation and defense \cite{ibitoye2019analyzing,akhtar2018threat}.

{To harden DNNs against adversarial attacks, there are two different approaches that have been explored. The first approach focuses on improving the classification accuracy of the DNN on adversarial inputs (adversarial accuracy) \cite{shafahi2019adversarial, madry2017towards, ganin2016domain}. This is achieved through techniques like adversarial training. In adversarial training, the DNN is trained on both clean and adversarial data till the DNN loss is minimized \cite{madry2017towards}. Other works have used image compression \cite{dziugaite2016study}, input randomization \cite{xie2017adversarial, xie2017mitigating} and parameter noise injection \cite{he2019parametric} to improve the adversarial accuracy of the DNN. The second approach is adversarial input detection. Here, the goal is to detect and reject adversarial inputs from propagating deeper into the network \cite{sterneck2021noise}.}

Most prior works on adversarial input detection augment neural network based detectors at intermediate layers \cite{metzen2017detecting, yin2019gat, sterneck2021noise}. Others use complex mathematical analysis to detect adversarial and clean activations \cite{grosse2017statistical}. These methods are 1) difficult to realise on hardware, and 2) vulnerable to adversarial attacks. For example, neural network-based detectors have been shown to be vulnerable to \textit{dynamic adversarial attacks} \cite{metzen2017detecting}. Here, the objective functions of both the neural network-based detector and the main DNN are used to generate much stronger attacks \cite{metzen2017detecting, sterneck2021noise}. To bypass these problems, we use a hardware driven approach. We monitor hardware signatures in DNN accelerators like currents in processing units to differentiate between adversarial and clean inputs. Several works on both digital von-neumann and analog crossbar based DNN accelerators have been proposed so far \cite{marinella2018multiscale, chen2018neurosim, shafiee2016isaac, chi2016prime, liu2016harmonica, xia2019memristive}. In this work, we focus on intercepting hardware signatures in analog crossbar based DNN accelerators. 
\begin{figure*}
    \centerline{\includegraphics[width=0.8\textwidth]{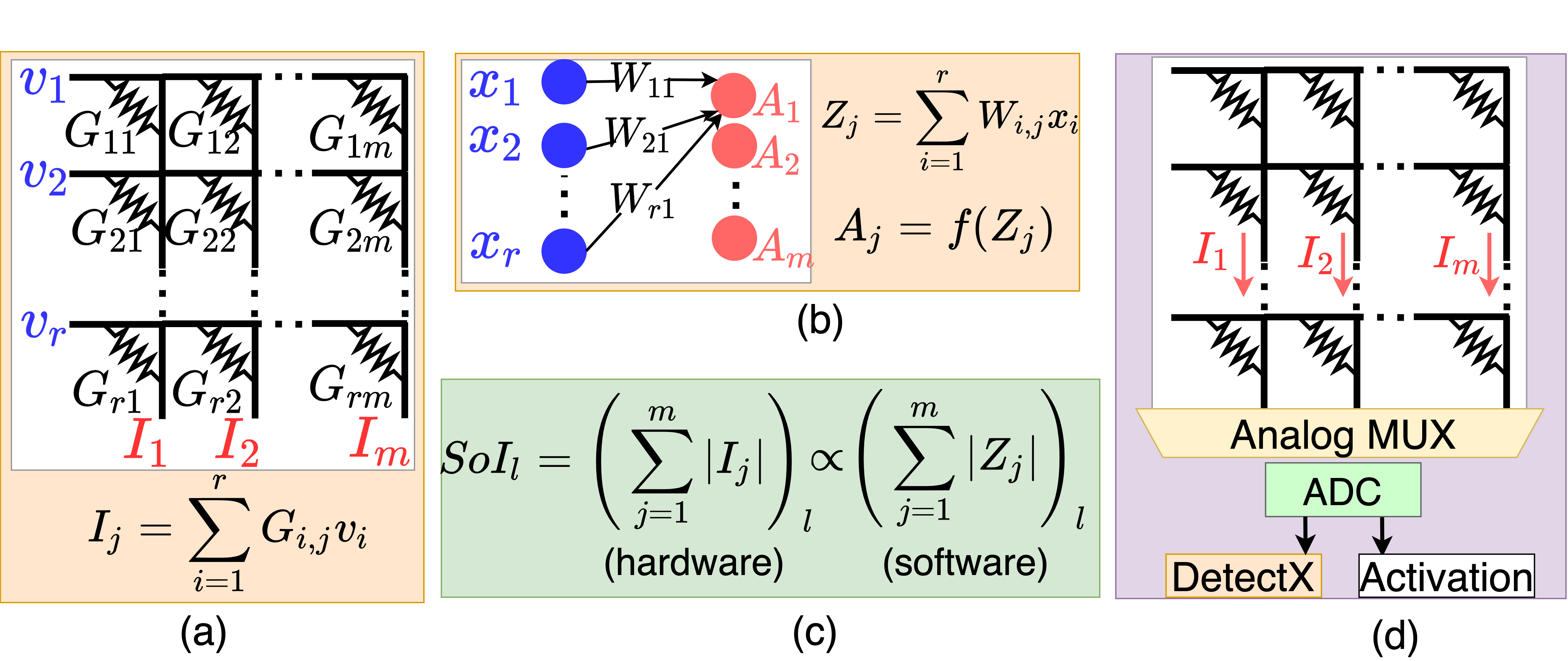}}
    \caption{a) MAC operations in analog crossbar arrays are realised using Ohm’s law and Kirchoff's Current Law resulting in output currents $I_j$. The current $I_j$ is the result of a weighted summation operation between conductances $G_{ij}$ and inputs $v_i$. b) In software, the MAC outputs $Z_j$ is realised using weighted summation operations between the weights $W_{ij}$ and inputs $x_i$. $A_j$ is the output of the activation function applied to $Z_j$ c) The SoI in analog crossbars can be computed by summing up the magnitudes of all the column currents ($I_i$ to $I_m$). This is proportional to performing absolute value summation of all the weighted summation outputs $Z_j$. d) The DetectX module is integrated with a Neurosim-like analog crossbar architecture. The column currents are multiplexed using an analog multiplexer which are then converted to digital values using the Analog to Digital Converter (ADC). For crossbars in the first layer, the output of ADC is fed into the DetectX module and the Activation circuits. However, for other layers, the DetectX module is absent.}
    \label{crossbar}
\end{figure*}

As seen in Fig. \ref{crossbar}a, in analog crossbars \cite{marinella2018multiscale, chen2018neurosim}, multiplication between input voltages $v_i$ and conductances $G_{ij}$ are realised using Ohm's law. The conductances are stored in Non Volatile Memory (NVM) devices like Resistive Random Access Memories (RRAMs) \cite{chen201865nm, wong2012metal}, Ferroelectric Field Effect Transistors (FeFETs) \cite{jerry2017ferroelectric,reis2018computing} among others. The accumulation is realised using summation of individual currents using Kirchoff's Current Law (KCL). This results in column currents $I_j$. Thus, $I_j$s are the result of a full Multiply Accumulate (MAC) operation. Equivalently, in software, MAC operations between inputs $x_1 - x_r$ and the weights $W_{ij}$ result in weighted summation values $Z{j}$. The activation value $A_j$ for each output neuron is obtained by applying an activation function $f$ on $Z_j$. This has been shown in Fig. \ref{crossbar}b.

In this work, we use the Sum of Currents (SoIs) parameter in analog crossbar arrays to detect adversarial inputs. For a particular layer $l$, SoI is the sum of the magnitudes of all the column currents in the crossbar as shown in Fig. \ref{crossbar}c. As an example, consider the crossbar shown in Fig. \ref{crossbar}a having column currents, $I_1 - I_m$. Here, the SoI value can be computed by summing up the magnitudes $I_1 - I_m$. From the software perspective, SoI is proportional to the sum of all the magnitudes of $Z_j$ in a particular layer $l$. Later, we will see that the SoIs from the first layer can be used to perform adversarial input detection. 

\begin{figure*}
    \centering
    \subfloat[][]{
    \includegraphics[width=0.33\textwidth]{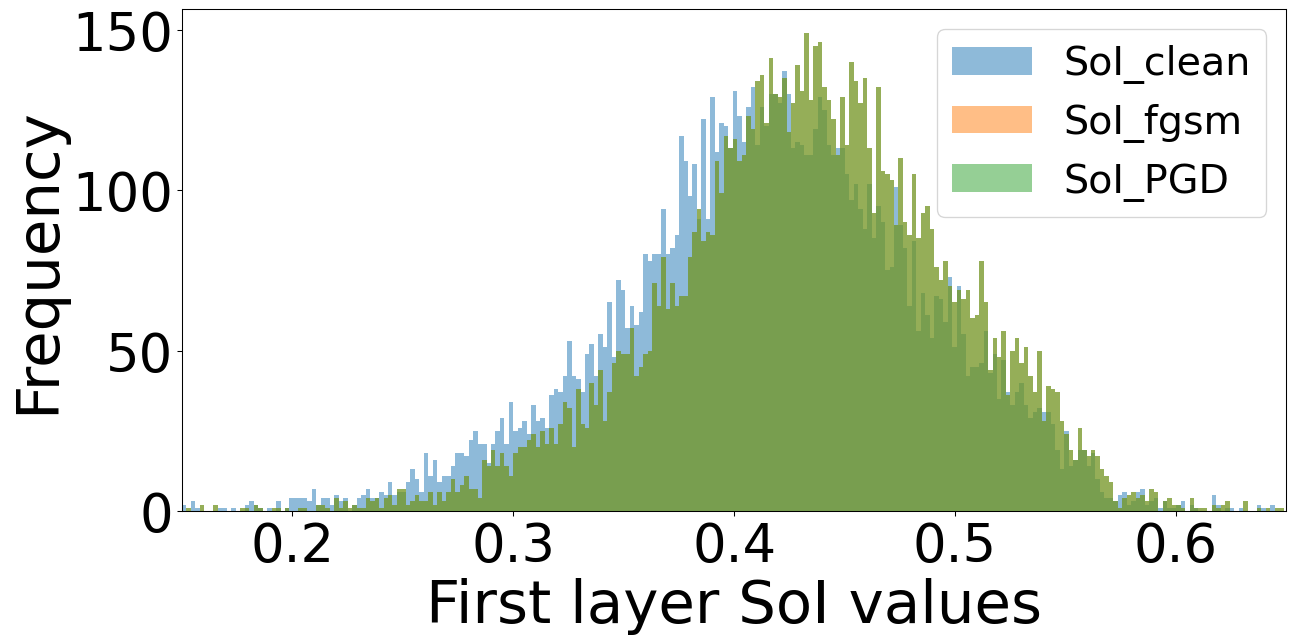}
    \label{soi_unseparated}}
    \subfloat[][]{
    \includegraphics[width=0.33\textwidth]{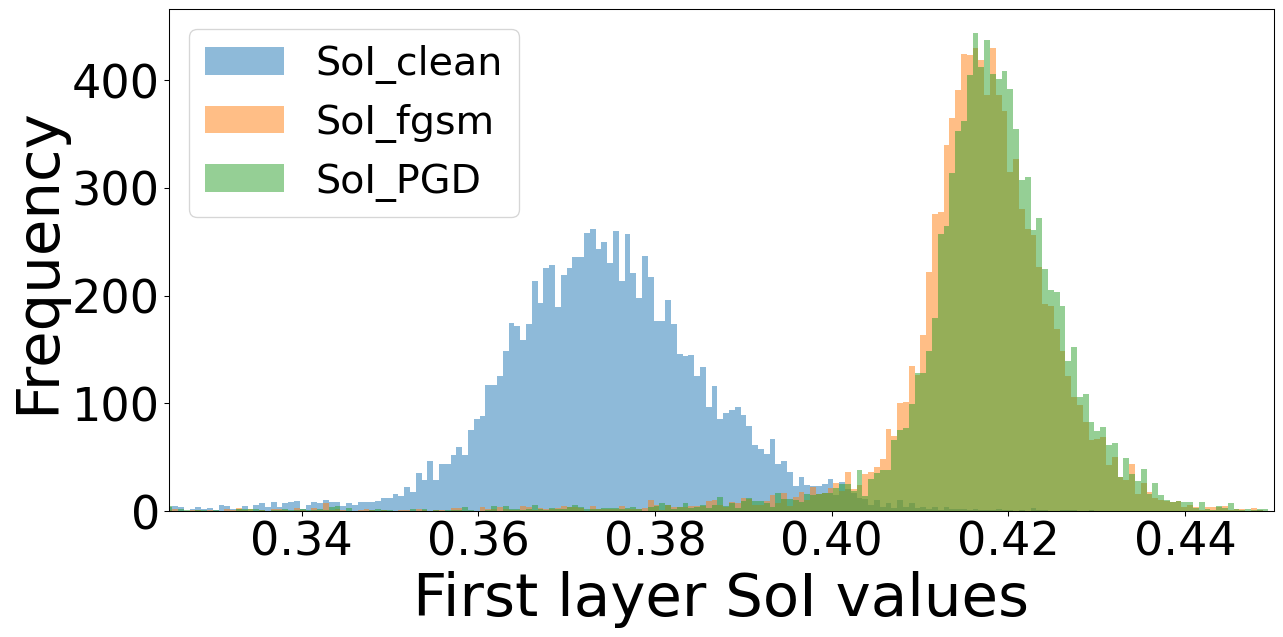}
    \label{soi_sep_sw}}
    \subfloat[][]{
    \includegraphics[width=0.33\textwidth]{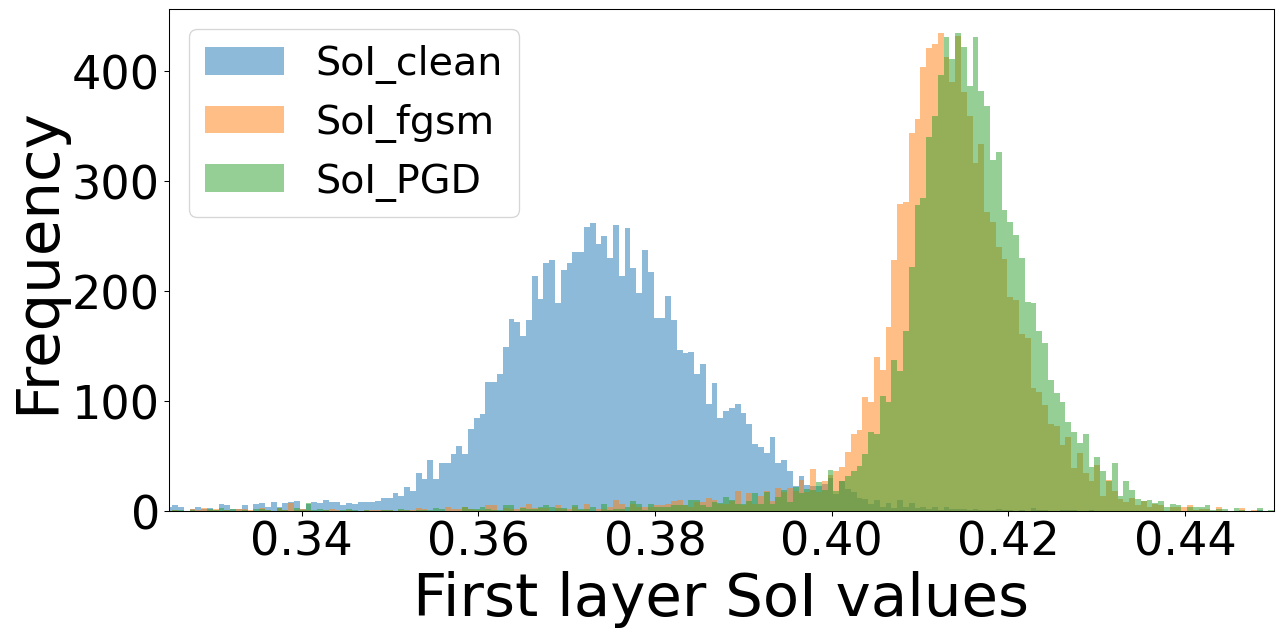}
    \label{soi_sep_hw}}
    \label{soi_before_after}
    \caption{Results of dual-phase training with CIFAR10 dataset on VGG8 network. a) The First layer SoI distribution corresponding to clean and adversarial inputs before the proposed dual-phase training methodology. The adversarial attacks correspond to PGD [$\epsilon$=16/255, $\alpha$=2/255, $n$=10] and FGSM [$\epsilon$=16/255] attacks b) Separated layer 1 clean and adversarial SoI distributions obtained after the dual-phase training. c) After implementing the dual-phase trained DNN on an ideal Neurosim-like crossbar architecture, we observe similar separation between clean and adversarial SoIs. In all the figures, the x-axis represents the range of SoI values and the y-axis represents the number of occurences of each SoI value (\textit{Frequency}).}
\end{figure*}

\textbf{Motivation:} Using SoI, we propose a detection mechanism that is hardware friendly and energy efficient. Absence of a neural network-based detector increases the robustness against \textit{dynamic adversarial attacks}. As seen in Fig. \ref{soi_unseparated}, the first layer adversarial SoIs are higher than clean SoIs. This is due to the added input perturbations. Note, the x-axis in Fig. \ref{soi_unseparated} depicts the range of SoI values and the term \textit{frequency} refers to the number of occurrences  of each SoI values. The adversarial SoIs are the SoIs corresponding to strong Projected Gradient Descent (PGD) and Fast Gradient Sign Method (FGSM) attacks. Likewise, clean SoIs are the SoIs corresponding to clean inputs. Clearly, in Fig. \ref{soi_unseparated}, the clean and adversarial SoI distributions are not well separated for reliable detection. 

To this end, we propose a dual-phase training methodology as follows: In the first phase, we train the DNN to increase the separation between first layer clean and adversarial SoIs. This is shown in Fig. \ref{soi_sep_sw}. When we deploy the dual-phase trained DNN on hardware, we see a similar separation between clean and adversarial SoIs as Fig. \ref{soi_sep_sw}. This is shown in Fig. \ref{soi_sep_hw}. This shows that the SoI separation in software leads to SoI separation in hardware. In the second phase, we freeze the first layer synaptic weights and explicitly perform adversarial training. This improves the accuracy of the DNN on clean inputs (clean accuracy). Additionally, it further improves the robustness of the DNN. After the dual-phase training, a Look Up Table (LUT) is created with entries as follows: The first column contains a list of sample SoI values. The second column consists of the probabilities associated with each SoI value. A high probability denotes that the SoI corresponds to a clean input. 

The dual-phase trained DNN is implemented on a Neurosim-like analog crossbar architecture \cite{peng2019dnn+}. For adversarial input detection, we integrate the analog crossbar with the DetectX module as shown in Fig. \ref{crossbar}d. The DetectX module is implemented using a 32nm CMOS Predictive Technology Model (PTM) \cite{balijepalli2007compact}. It contains a \textit{SoI Computing Unit} and a \textit{LUT-based detector}. During detection, the SoI value computed by the \textit{SoI Computing Unit} is looked up in the LUT and a confidence score denoting the probability of clean input is generated. If the confidence score is low, the input is treated as adversarial and rejected. For hardware evaluation of the analog crossbar+DetectX system, we use Neurosim platform. Additionally, we perform robustness evaluation using different datasets like CIFAR10, CIFAR100 \cite{krizhevsky2009learning} and TinyImagenet \cite{le2015tiny}. 

To the best of our knowledge, this is the first work that exploits hardware based signatures for adversarial input detection in analog crossbar arrays. In summary, the key contributions of our work are as follows.:
\begin{enumerate}
    \item We propose DetectX, a digital CMOS-based adversarial input detector. DetectX is a small digital module that uses SoI values in analog crossbar architectures to detect adversarial attacks. For this, it is appended at the end of an analog crossbar array. Note, although DetectX is a digital module, the crossbar+DetectX system is a mixed system. 
    \item To improve the SoI based detection performance, we propose a dual-phase training algorithm. The first phase is geared towards increasing the SoI separation between clean and adversarial inputs. While the second phase improves the adversarial robustness of the DNN on adversarial inputs that are not rejected by the detector. 
    \item We validate our methodology through experiments performed on the Neurosim platform. We use different datasets like CIFAR10, CIFAR100, and TinyImagenet. Additionally, we conduct a comprehensive analysis to compare the energy-robustness tradeoffs between DetectX and other state-of-the-art adversarial input detection works. DetectX outperforms these works in energy efficiency with comparable robustness against different adversarial attacks. Additionally, we also show that DetectX can improve the robustness of models against Black-Box and \textit{Dynamic Adversarial Attacks}.
\end{enumerate}

\section{Background}
\subsection{Analog Crossbar Arrays}
Analog crossbar arrays have recently emerged as an energy efficient method for implementing MAC operations. In an analog crossbar, multiplications between inputs and synaptic weights are realised using Ohm's law. As shown in Fig. \ref{crossbar}a, the DNN weights are stored as conductance states $G_{ij}$ of NVM devices (RRAMs,FeFETs) \cite{jerry2017ferroelectric,chen201865nm}. The largest synaptic weight is mapped to the highest conductance state and vice-versa. Likewise, the inputs are encoded as voltage values $v_i$. When a voltage is applied across an NVM device, a current proportional to $G_{ij}$ and $v_i$ is generated at a cross-point. The final MAC output is obtained by adding the individual device currents along a column by the virtue of KCL.

\subsubsection{The Neurosim Platform}
Neurosim \cite{chen2018neurosim} is a Python based platform that performs a holistic energy-latency-accuracy evaluation of analog crossbar-based DNN accelerators. The Neurosim platform supports both SRAM and memristive-based (RRAM and FeFET) computing devices. 

\textbf{Neurosim-like analog crossbar architecture:} In this work, we use a Neurosim-like analog crossbar architecture as shown in Fig. \ref{crossbar}d for all hardware evaluations. In the Neurosim-like analog crossbar, multiple columns are multiplexed using an analog multiplexer. The multiplexer serializes parallel input currents for the analog to digital conversion stage. This essentially reduces the number of Analog to Digital Converters (ADC) required at the cost of increased computation time (latency). The ADC converts analog currents to digital values using current-to-voltage sense amplifiers \cite{jiang2021analog}. After the ADC stage, the activation unit applies an activation function like ReLU \cite{arora2016understanding} or sigmoid \cite{zadeh2010daily} to the digital MAC outputs. Note, digital implementation of mathematical functions are more stable and hardware efficient than analog implementations. 

For implementing DNNs on the Neurosim platform, first several hardware parameters are initialized. These include the crossbar size, on-off ratio of the memristive device, the conductance variation parameter and so on. The DNN is first mapped onto several analog crossbars using Neurosim's novel, energy-efficient mapping strategy. The mapping strategy reduces the number of memory accesses during the MAC operations. After the DNN mapping stage, the Neurosim platform performs feed-forward operations on the inputs to the DNN. During this, MAC operations between the inputs and the DNN weights are performed in the crossbars resulting in output currents. The output currents are then converted to digital values by the ADC. During the MAC operations, Neurosim incorporates hardware non-idealities like conductance variation effects in memristive devices and data-quantization loss. This facilitates a realistic hardware evaluation of DNNs. Additionally, the Neurosim platform also computes the energy and latency of different hardware components used during the feed-forward operation. These include the analog crossbars and the peripheral circuits like the MUX, ADC, activation units among others.

Although, the Neurosim platform incorporates various hardware-based non-ideal effects like quantization noise and memristive conductance variations, it does not incorporate the non-idealities arising due to process-voltage-temperature variations (PVT), device to device variations and so on. 

In this work, for hardware based adversarial input detection, we append the DetectX module after the ADC stage of the first layer crossbar. The DetectX module uses the SoI parameter for adversarial input detection. It consists of a \textit{SoI computing unit} for SoI value computation and a \textit{LUT-based detector} that generates the detector output. 

\subsection{Adversarial attacks}
In this paper we discuss two kinds of adversarial attacks: White-Box (WB) attacks and Black-Box (BB) attacks. In WB attacks, the attacker has complete access to the DNN network and parameters. This is a more difficult and stronger adversarial attack. BB attacks are easier to execute as they do not require complete access to the target DNN. The adversarial inputs can be generated using a different DNN model. In this work, we discuss two gradient based adversarial attacks. 
\begin{enumerate}
\item \textit{Fast Gradient Sign Method (FGSM)} \cite{kurakin2016adversarial} is a one step gradient based attack shown in Equation \ref{fgsm}. First, the gradients of the DNN loss $\mathcal{L}(\theta,x,y_{true})$ with respect to the input $x$ are calculated. Then, a $sign()$ operation converts the gradients into unit directional vectors. The unit vector is multiplied by a scalar perturbation value, $\epsilon$. Finally, the perturbation vector is added to the input $x$ to create an adversarial data. Note, that perturbations are added to $x$ along the direction of the gradients to maximize DNN loss $\mathcal{L}$. 
\begin{equation}
    x_{adv} = x + \epsilon ~sign(\nabla_x(\mathcal{L}(\theta,x,y_{true})))
    \label{fgsm}
\end{equation}

\item {\it Projected Gradient Descent (PGD)} attacks have been shown to cause highly effective adversarial attacks \cite{madry2017towards}. The PGD attack, shown in Equation \ref{pgd} is an iterative attack over $n$ steps. In each step $i$, perturbations of strength $\alpha$ are added to $x_{adv}^{i-1}$. Note, that $x_{adv}^{0}$ is created by adding random noise to the clean input $x$. Additionally, for each step, $x_{adv}^{i}$ is projected on a \textit{Norm ball} \cite{madry2017towards}, of radius $\epsilon$. The type of \textit{Norm ball} ($L_{\infty}, L_{2}$ and so on) used signifies how the net perturbations with respect to the input is computed. In this work, we use the $L_{\infty}$ \textit{Norm ball} (of radius $\epsilon$) projection for all the PGD attacks. In other words, we ensure that the maximum pixel difference between the clean and adversarial inputs is $\epsilon$. 
\begin{equation}
    x_{adv} = \sum_{i=1}^{n} x_{adv}^{i-1}+\alpha ~sign(\nabla_x\mathcal{L}(\theta, x, y_{true}))
    \label{pgd}
\end{equation}
\end{enumerate}
\textbf{Strength of an Adversarial Attack: }
The strength of an adversarial attack is directly proportional to the amount of perturbations added to the input. To create strong adversarial attacks, high values of $\epsilon$ (in case of FGSM attacks) and $\alpha, \epsilon, n$ (for PGD attacks) are chosen. Similarly, weak adversarial attacks are created by choosing extremely small values of $\alpha, \epsilon, n$. For moderate strength attacks, the values of $\alpha, \epsilon, n$ lie in between the weak and strong adversarial attacks. 

\subsection{Performance Metrics for Adversarial Input Detection}
\label{auc_error_accuracy}
To evaluate the adversarial detection performance, we use three performance metrics: \textit{ROC-AUC, Accuracy} and \textit{Error}. 
\begin{itemize}
    \item \textbf{Area Under the ROC Curve (ROC-AUC):} In this work, we use ROC-AUC score to analyse the performance of DetectX. The ROC-AUC for a detector is high if the detector has a high confidence score for positive classes and low confidence score for the negative classes. Previous works have shown that using ROC-AUC is more stable than True Positive Rate (TPR) and False Positive Rate (FPR) based performance analysis \cite{yin2019gat}.  
    \item \textbf{Accuracy and Error:} In the context of adversarial input detection, {\it{Accuracy}} and {\it{Error}} are not complements of each other.  {\it{Accuracy}} refers to the fraction of clean inputs that are correctly classified and not rejected by the adversarial input detector. While, {\it{Error}} refers to the fraction of adversarial inputs that are classified incorrectly and not rejected by the adversary detector.
    
\end{itemize}
\section{Related Works}
\subsection{Works improving adversarial robustness of DNNs}
\subsubsection{Algorithmic approaches} Among various algorithmic works that try to improve the adversarial accuracy of the DNN, adversarial training has been shown to achieve state-of-the-art performance \cite{madry2017towards}. Adversarial training aims to solve the min-max problem. This is done by first creating a suitable adversarial dataset such that the DNN loss is maximized. Next, the DNN is trained on the adversarial dataset until the loss value converges to a minimum. Additionally, adversarial training has been shown to posses higher transferability across different types of attacks. Further, He et al. \cite{he2019parametric} showed that adding noise into DNN parameters during adversarial training can improve the adversarial robustness of the DNN. Other works by Xie et al. \cite{xie2017mitigating} and Dziugaite et al. \cite{dziugaite2016study} show that input level randomization and compression, respectively, can improve the adversarial robustness of the DNN. However, these works do not use an adversarial training approach. The review work by \cite{MIT_survey} provides a comprehensive list of different algorithm level DNN robustness improvement methodologies.
\subsubsection{Hardware approaches}
Recently hardware centred approaches have been looked at from the perspective of improving the adversarial robustness of DNNs. The work by Panda et al., QUANOS \cite{quanos}, leverages data quantization in order to minimize the adversarial noise sensitivity parameter. The work by Lin et al., Defensive Quantization, \cite{dq} used quantization in order to minimize the adversarial noise magnification effect in the deeper layers. This is done by optimizing the DNN to have a lipschitz constant less than 1. Addtionally, works like \cite{moitra2020exposing, bhattacharjee2020rethinking} show that intrinsic bit-error noise in 6T SRAM cells in hybrid CMOS memories and non-idealities in analog crossbar arrays can improve the adversarial robustness of DNNs.  
\subsection{Works based on adversarial input detection}
Towards adversarial input detection, most works so far have focused on algorithmic approaches. wherein the DNN is either trained on augmented data or the network architecture of the DNN is modified in order to facilitate adversarial input detection. The work by Grosse et al. augmented the dataset with an additional class representing adversarial inputs. The adversarial inputs were classified at the end of the DNN \cite{grosse2017detection, gong2017adversarial}. In contrast to this, Feinman et al. used a linear classifier at the end of the last hidden layer to classify the points lying away from the data manifolds as adversarial inputs \cite{feinman2017detecting}. The work by Yin et al. proposed training a binary classifier on subspaces created by partitioning the input space \cite{yin2019gat}. This resulted in multiple detectors that were used for adversarial input detection. Additionally, the work Li and Li proposes to detect adversarial features by training cascades of SVM classifiers on the Principal Component Analysis (PCA) of the outputs of the convolution layers of the main network \cite{li2017adversarial}. 

Additionally, Metzen et al. and Sterneck et al. appended a binary classifier between convolutional layers to detect adversarial inputs using the intermediate activation maps as features \cite{metzen2017detecting, sterneck2021noise}. While Metzen et al. used a heuristic approach to append adversarial input detectors at the end of intermediate convolutional layers, Sterneck et al. strategically placed the detector at the end of a convolution layer chosen using a metric called the adversarial noise sensitivity. This improved the energy and resource efficiency of the detection. However, both the works showed that their detector performances are affected by dynamic adversarial attacks.  

In this work, we take a hardware based approach towards adversarial input detection. Using DetectX, we show that the parameter SoI can be used to perform energy efficient adversarial input detection in analog crossbar based DNN accelerators. Apart from energy efficiency, DetectX also improves the overall adversarial robustness of the DNN against black-box, white-box and dynamic adversarial attacks when compared to previous works. 

\section{Dual-Phase Training Methodology }
\begin{algorithm}
\SetAlgoLined
\textbf{Training Phase 1}\\
\textbf{Inputs:} Pre-trained DNN model; Training dataset: 1:1 ratio of clean and adversarial data\;
\For{epochs $\le$ num\_epochs}{
Calculate first layer SoI\;
Forward pass\;
Compute loss using Equation \ref{loss}\;
Backward pass\;
}
{\textbf{Training Phase 2}\\
\textbf{Inputs:} DNN after Training Phase 1;
Training dataset: 1:1 ratio of clean and weak adversarial data\;
\For{epochs $\le$ num\_epochs}
{
Forward pass\;
Compute loss using Equation \ref{ce_loss}\;
Backward pass\;}}

\caption{Proposed Dual-Phase Training Methodology}
\label{proposed_methodology}
\end{algorithm}
The dual-phase training methodology is divided into two phases. The first phase is geared towards increasing the SoI separation between clean and adversarial SoIs. In the second phase, we perform adversarial training to further improve the adversarial robustness of the DNN.

\subsection{Phase1 Training}
\label{Training Phase 1}
In the first phase training, a pre-trained model (DNN trained on clean data in the standard manner) is trained on dataset consisting of [clean+strong adversarial data]. For this we use the objective function shown in Equation \ref{loss}. The adversarial data is generated using PGD attacks on the pre-trained DNN. The values $\lambda_c$ and $\lambda_a$ represent the desired clean and adversarial SoI values. The mean square loss ($\mathcal{L_{MSE}}$) represents the Eucledian distances between the actual SoI values and the desired SoI values. During training, for adversarial inputs, $y$ activates the term $\mathcal{L}_{MSE}(SoI_a,\lambda_a)$ while deactivating the term $\mathcal{L}_{MSE}(SoI_c,\lambda_c)$ simultaneously and vice versa. We take y=1 when the input is adversarial and y=0 when the input is clean. For better convergence of Equation \ref{loss}, the cross-entropy loss term $\mathcal{L}_{CE}$ is scaled by a small factor $\beta$ (of the order of $10^-6$. 
\begin{multline}
    \mathcal{L} = \beta ~\mathcal{L}_{CE}+y ~\mathcal{L}_{MSE}(SoI_{a}, \lambda_{a})\\
    +~(1-y) ~\mathcal{L}_{MSE}(SoI_{c}, \lambda_{c})
    \label{loss}
\end{multline}
\begin{figure}
\centerline{\includegraphics[width=0.4\textwidth]{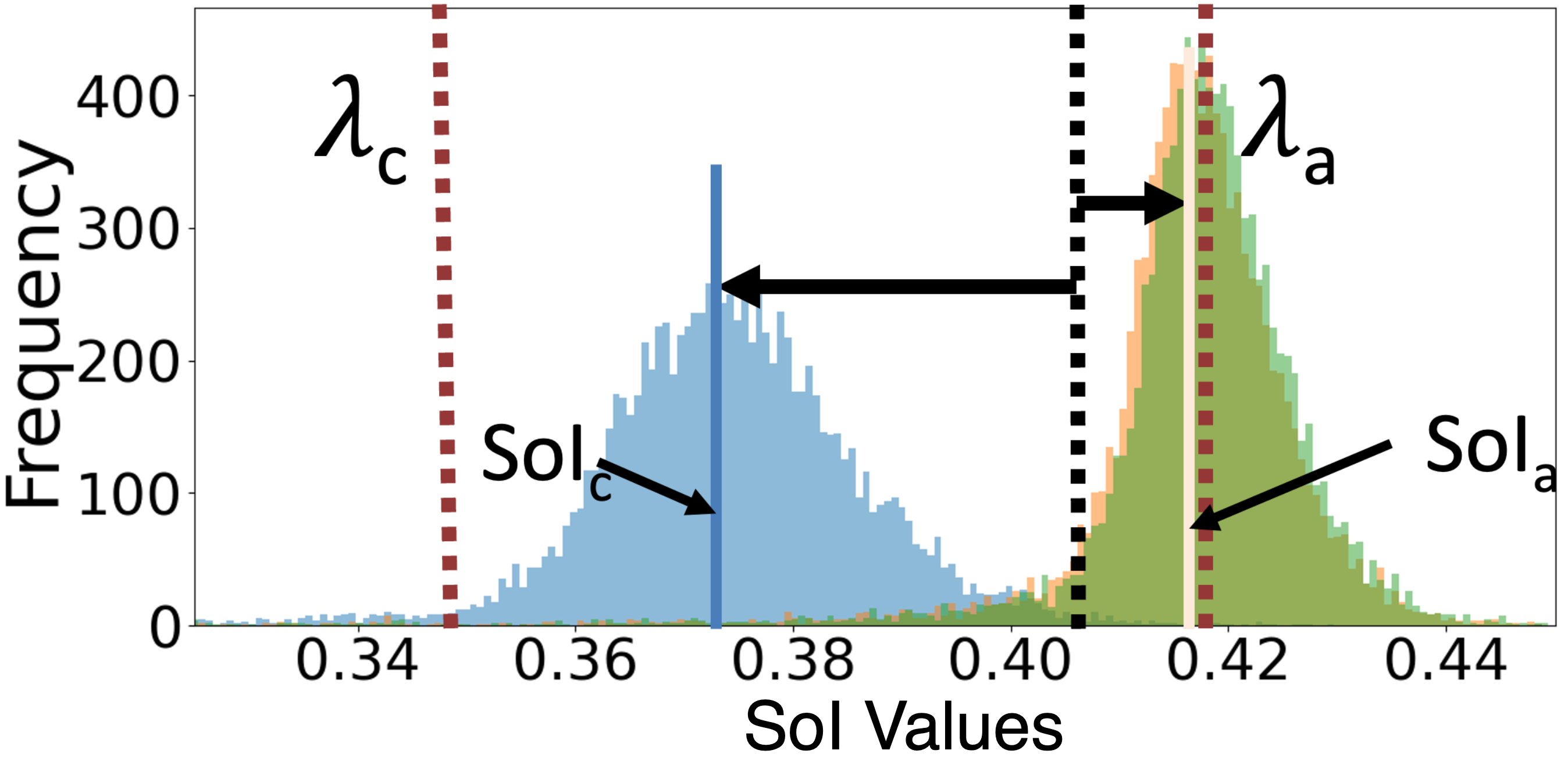}}
\caption{After Phase1 training, the mean SoI values corresponding to clean and adversarial inputs ($SoI_c$ and $SoI_a$) are well separated. Additionally, they are shifted closer to the desired SoI values ($\lambda_c$ and $\lambda_a$). This shift is shown using the horizontal arrows. The black dotted line represents the SoI of clean and adversarial inputs before the Phase 1 training. The SoIs correspond to the CIFAR10 dataset on a VGG8 network.}
\label{explain_sois}
\end{figure}
Fig. \ref{explain_sois} demonstrates the working of Phase1 training. Before phase 1 training, the mean SoI values for clean and adversarial inputs (black dotted line) are almost indistinguishable. Consequently, the Eucledian distances ($SoI_a$, $\lambda_a$) and ($SoI_c$, $\lambda_c$) are very large. During training, the $SoI_a$ and $SoI_c$ move towards $\lambda_a$ and $\lambda_c$ respectively (shown using the horizontal arrows). This leads to reduction in the net Eucledian distances and increases the SoI separation. 

During Phase1 training the weights of the first layer are optimized for increased SoI separation. Additionally, the down scaling of the term $\mathcal{L}_{CE}$ in Equation \ref{loss} compromises the \textit{clean accuracy} of the DNN. Moreover, as we will discuss later, the performance of the detection performance of DetectX is slightly lowered for extremely weak adversarial attacks (i.e, adversarial attacks having small input perturbations). Thus, in order to improve the \textit{clean accuracy} and further improve the robustness of the DNN, we perform Phase2 training.  
\subsection{Phase2 Training}
In the second phase, we explicitly perform adversarial training on the DNN using the objective function shown in Equation \ref{ce_loss}. The adversarial training set comprises of [clean + weak adversarial data]. During adversarial training, we freeze the first layer weights in order to maintain the separation between clean and adversarial SoIs. 
\begin{equation}
    \mathcal{L} = \mathcal{L}_{CE}
    \label{ce_loss}
\end{equation}

\subsection{Creating the SoI-Probability LUT}
\label{LUT create}
\begin{figure}
\centerline{\includegraphics[width=0.4\textwidth]{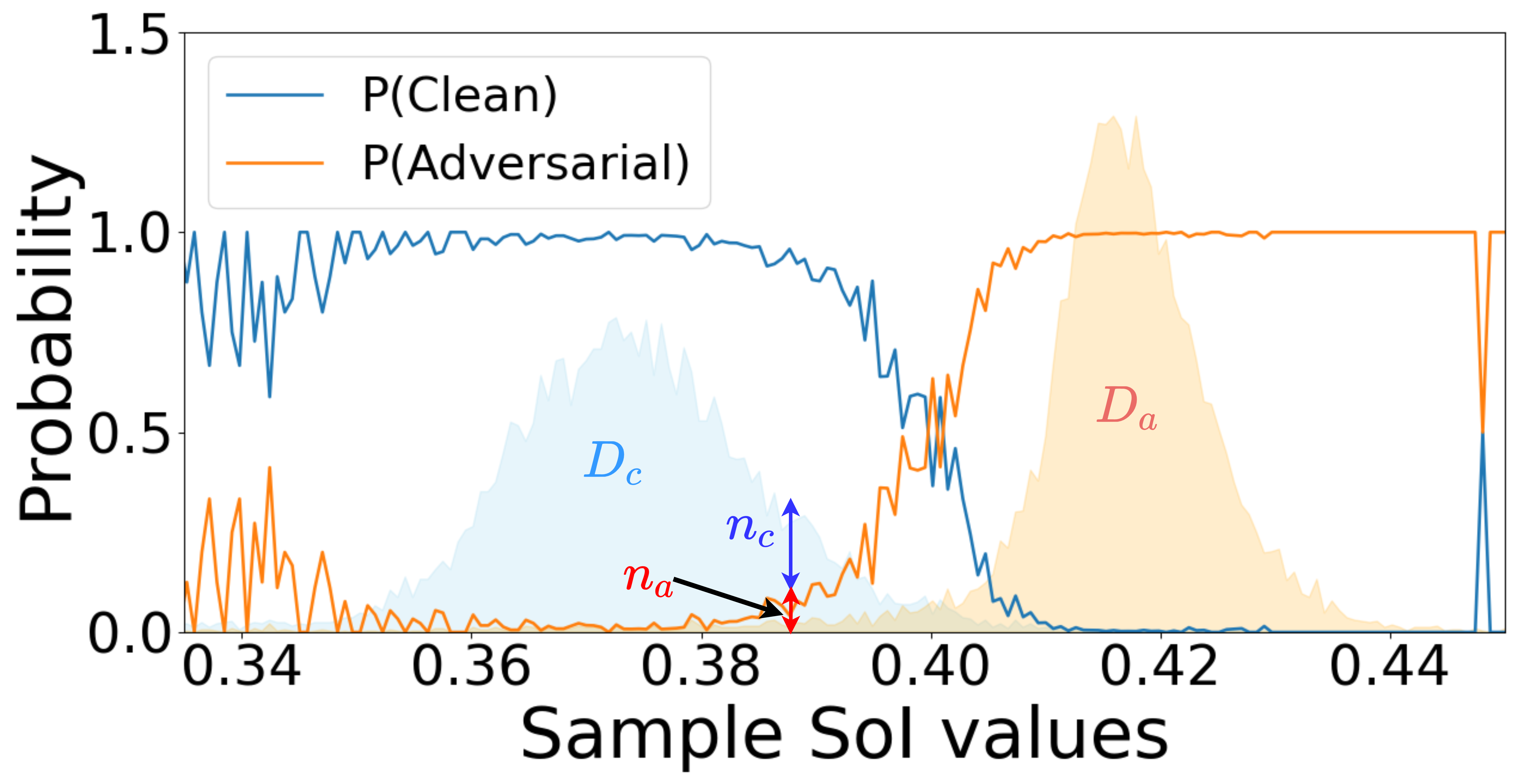}}
\caption{After Training Phase 1, SoIs $D_c$ corresponding to clean (blue) and $D_a$ corresponding to adversarial (orange) inputs are well separated which helps in improved adversarial detection. The clean and adversarial histogram is used to create the probability LUT stores the sample SoI values in the first column and the corresponding P(clean) values in the second column. Note, here we demonstrate SoI-Probability LUT generation for the CIFAR10 dataset on a VGG8 network.}
\label{prob_list}
\end{figure}

The SoI-Probability LUT has the following structure: The first column stores the sample SoI values while the second column stores the corresponding clean probabilities $P(Clean)$ for each SoI value. Fig. \ref{prob_list} shows the plot of the $P(Clean)$ and $P(Adversarial)$ values against the sample SoI values. The value of $P(Clean)$ denotes the probability that a given SoI value belongs to a clean input. It is calculated using Equation \ref{clean}. Here, $n_a$ and $n_c$ denote the frequencies of adversarial and clean SoIs, respectively, at a particular sample SoI value. Note, $P(Adversarial)$ (shown in orange), is not stored in the LUT. For creating the SoI-Probability LUT, we select a random sample of clean images. To create the adversarial dataset, we add adversarial perturbations to the randomly selected clean images. Then, the SoI values are computed for both the clean and adversarial images. In this work, we refer to $D_c$ as the SoI distribution corresponding to clean images and $D_a$ as the SoI distribution corresponding to adversarial images. The choice of the adversarial distribution $D_a$ is critical to the detector performance and will be discussed in the later sections. 
\begin{equation}
    P(Clean) = \frac{n_{c}}{n_{c}+n_{a}}
    \label{clean}
\end{equation}

\subsection{Why the Dual-Phase Training Works?}
\label{Why the Dual-Phase Training Works?}
From Fig. \ref{crossbar}c, SoIs can be expressed as the absolute value of all the weighted summation outputs in a particular layer of the DNN. Moreover, in the first layer, the convolution operation is merely a linear operation on the inputs. In other words, any change in the input will affect the output linearly. We use these properties to understand why the dual-phase training works. 

From Fig. \ref{crossbar}b and Fig. \ref{crossbar}c, SoI for clean input data, $X_{clean}$, can be written as shown in Equation \ref{soi_clean}. For creating adversarial inputs $X_{adv}$, perturbations $\mathcal{P}$ are added to $X_{clean}$ as seen in Equation \ref{adv_image}. These added perturbations lead to a significant change in the adversarial SoI value as seen in Equation \ref{soi_adv}.
\begin{equation}
    SoI_{clean} = \sum_{j=1}^{m}{\sum_{i=1}^{r}} |W_{ij}X_{clean,i}|
    \label{soi_clean}
\end{equation}
\begin{equation}
     X_{adv} = X_{clean} + \mathcal{P}
     \label{adv_image}
\end{equation}
\begin{multline}    
SoI_{adv} = \sum_{j=1}^{m} {\sum_{i=1}^{r}} |W_{ij}(X_{clean,i}+\mathcal{P}_i)|\\
= \sum_{j=1}^{m} {\sum_{i=1}^{r}} |W_{ij}X_{clean,i}+ W_{ij} \mathcal{P}_i)|
\label{soi_adv}
\end{multline}

\subsection{Evaluating SoI separation} 
\label{Evaluating SoI separation}
\begin{figure*}
    \centering
    \includegraphics[width=0.8\textwidth]{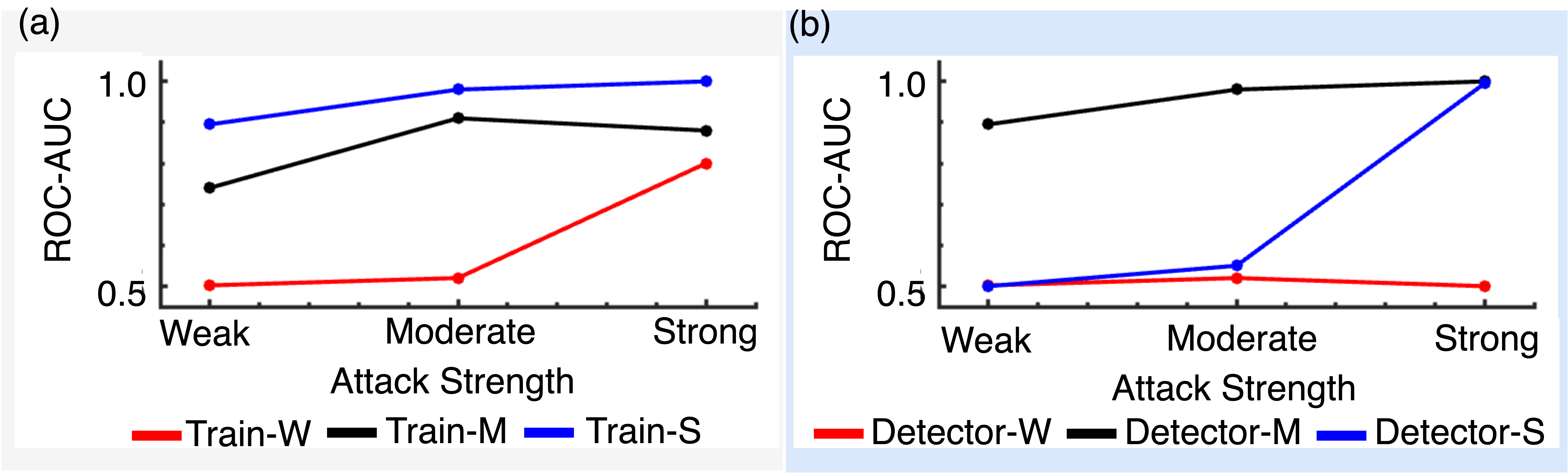}
    \caption{In a) strong, moderate and weak attacks were used in the Phase1 training. These scenarios are denoted by Train-S, Train-M and Train-W respectively. Training with strong adversarial attacks leads to higher detection performance across weak, moderate and strong attacks. In b) The probability LUT is generated using $D_c$= clean and $D_a$ = Weak (Detector-W), $D_a$ = Moderate (Detector-M) and $D_a$ = Strong (Detector-S). Notice, here that generating probability LUT using moderate strength attacks allow better performance against different strengths of attacks. Although, we show the results obtained for CIFAR10 dataset on a VGG8 network, the trends of the ROC-AUC scores remain constant across different datasets.}
    \label{auc_vary}
\end{figure*}

In this section, we will evaluate the Phase1 training performance based on different strengths of adversarial data. For this, a DNN is trained using the CIFAR10 dataset comprising of [clean+weak/ moderate/ strong adversarial data]. Fig. \ref{auc_vary}a shows the ROC-AUC scores corresponding to weak (\textit{Train-W}), moderate (\textit{Train-M}) and strong (\textit{Train-S}) attacks under three different scenarios. In {\it Train-W}, the DNN is trained with weak adversarial data. Here, the ROC-AUC scores across all strengths of adversarial attacks are low (small separation between clean and adversarial SoIs). Although, with moderate strength adversarial attacks, the performance is better than the previous case, it is not able to achieve a higher SoI separation for strong attacks. Finally, in the case of \textit{Train-S}, training with strong adversarial data achieves the highest SoI separation across different attack strengths. This is signified by an overall high ROC-AUC score across all attack strengths. 

\subsection{Choosing adversarial distribution $D_a$ for SoI-Probability LUT creation} In this section, we analyse the performance of DetectX based on three different adversarial SoI-Probability LUTs. For each LUT, the clean SoI distribution $D_c$ is fixed while adversarial SoIs $D_a$ can be different depending on the type of adversarial attack used.

We use a dual-phase trained DNN (trained on the CIFAR10 dataset [Clean+Strong PGD data]). This leads to maximum SoI separation as discussed in Section \ref{Evaluating SoI separation}. Then, we create three different SoI-Probability LUTs with $D_a$ = weak ({\it Detector-W}), $D_a$ = moderate ({\it Detector-M}) and $D_a$ = strong ({\it Detector-S}) adversarial SoI distributions. Fig. \ref{auc_vary}b shows the ROC-AUC scores corresponding to different strengths of adversarial attacks for the three different SoI-Probability LUTs. While {\it Detector-W} performs poorly across all strengths of attacks, {\it Detector-S} performs well with strong attacks only. Here, {\it Detector-M} has a high performance across all strengths of adversarial attacks. Thus, SoI-Probability LUT created using $D_a$ corresponding to moderate strength attacks are unbiased and will be used in all further experiments in this paper. 

\section{Hardware Implementation}
In this section, we first define the hardware architecture of the DetectX module. Then we integrate it with an analog crossbar array to perform hardware-based adversarial input detection. 

\subsection{DetectX}
\begin{figure*}
    \centerline{\includegraphics[width=0.9\textwidth]{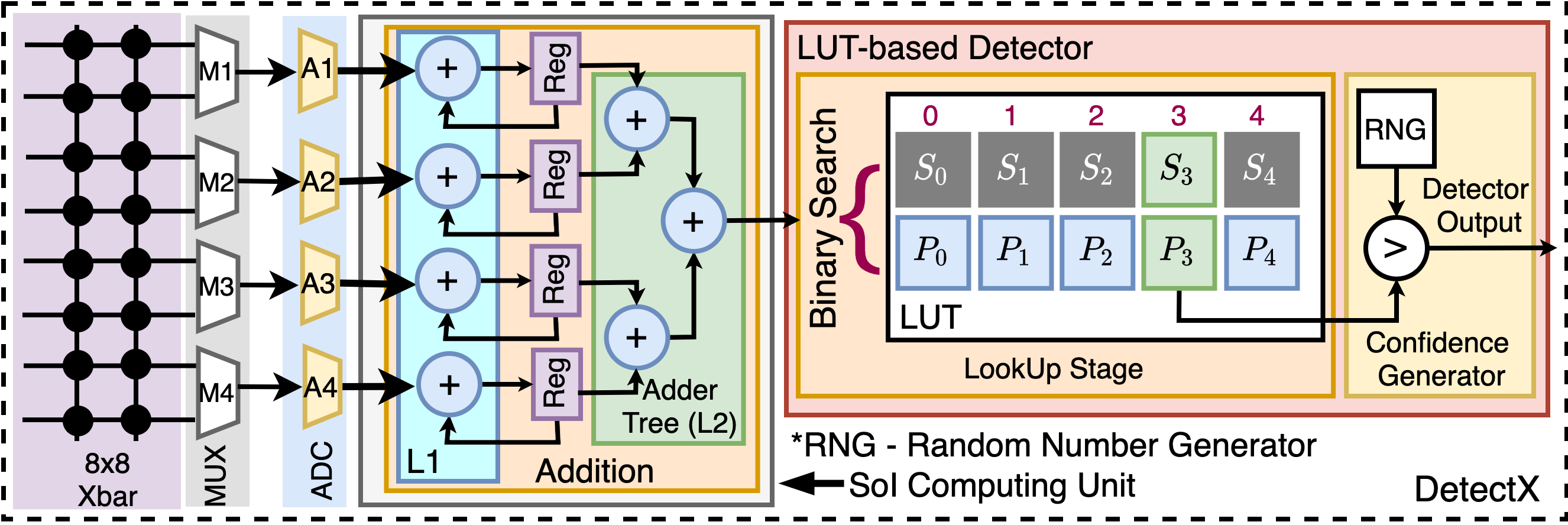}}
    \caption{The DetectX module receives MAC outputs from the ADC stage (A1-A4) of the crossbar (last two rows of an 8x8 crossbar shown for example). The \textit{L1 Adders} can support addition/subtraction depending on the most significant bit (MSB) value. This facilitates absolute value summation of the MAC values. Using registers \textit{Reg} allows addition of MAC values from cycle to cycle. After all the cycles are completed, the \textit{L2 Adders} compute the final SoI values which are passed to the \textit{LUT-based Detector}. Here, a \textit{Binary Search} is performed on the LUT and a sample SoI value $S_k$ and its corresponding P(clean) value $P_k$ is selected depending on the SoI value. The $P_k$ is then compared with a random sample generated by the $RNG$ in the \textit{Confidence Generator}. If the sample is less than $P_k$, the detector output is 1, meaning that the input is clean and vice-versa}
    \label{detectX}
\end{figure*}
As shown in Fig. \ref{detectX}, DetectX consists of two sub-modules: 1) The \textit{SoI Computing Unit}, and 2) The \textit{LUT-based detector}. To show how the DetectX module is integrated with an analog crossbar, we show an 8x8 crossbar having four 2:1 multiplexers. The outputs of the four multiplexers are fed as inputs to four ADCs A1-A4. The outputs of the A1-A4 are the inputs to the \textit{L1 Adders}. The \textit{L1 Adders} can support both addition and subtraction operations. If the most significant bit of the input is 1, then a subtraction operation is performed and addition otherwise. With this, we implement the absolute value summation operation for computing SoIs. To implement an accumulation operation, we use registers \textit{Reg}. Each \textit{Reg} stores the intermediate summation outputs. At the end of all the read cycles, the final summation value is forwarded to the \textit{L2 Adders}. As an example, in Fig. \ref{detectX}, due to the presence of four 2:1 MUXs, a total of 2 read cycles are required to read all the outputs from the crossbar. In each cycle, the analog current outputs are converted to digital values. These digital inputs are then added/subtracted to the previous accumulated values stored in the registers using the {\it L1 Adders}. After completion of all the read cycles, the accumulated values in each \textit{Reg} are added using {\it L2 adders} in the adder-tree. This gives the final SoI value. The size of the \textit{SoI Computing Unit} depends on the crossbar parameters (see Section \ref{Integration and Hardware Evaluation}). 

Next, the computed SoI value is passed to the \textit{LUT-based detector}. Here, the SoI value is looked up (\textit{LookUp Stage}) in the \textit{LUT} memory \cite{marinella2018multiscale}. The \textit{LUT} stores the sample SoI values ($S_0$ to $S_n$) and the corresponding $P(Clean)$ values $P_0$ to $P_n$. It is composed of 6T-SRAM cells. In the \textit{LookUp Stage}, a binary search operation is performed. Depending on the value of the input SoI, the $P_k$ corresponding to $S_k$ (such that $S_k$ $<$ SoI and $S_{k+1}$ $>$ SoI) is chosen and passed to the \textit{Confidence Generator}. The \textit{Confidence Generator} generates a random sample using a SRAM-based Random Number Generator(RNG) \cite{clark2018sram}. The RNG exploits the principle of bit-instability of SRAM cells during the turn-on phase. The RNG output is then compared with the input probability value $P_k$. If the RNG output is lesser than $P_k$, then the \textit{Detector Output} is 1. This means that the SoI belongs to a clean input and vice-versa.  
\begin{figure}
    \centerline{\includegraphics[width=0.5\textwidth]{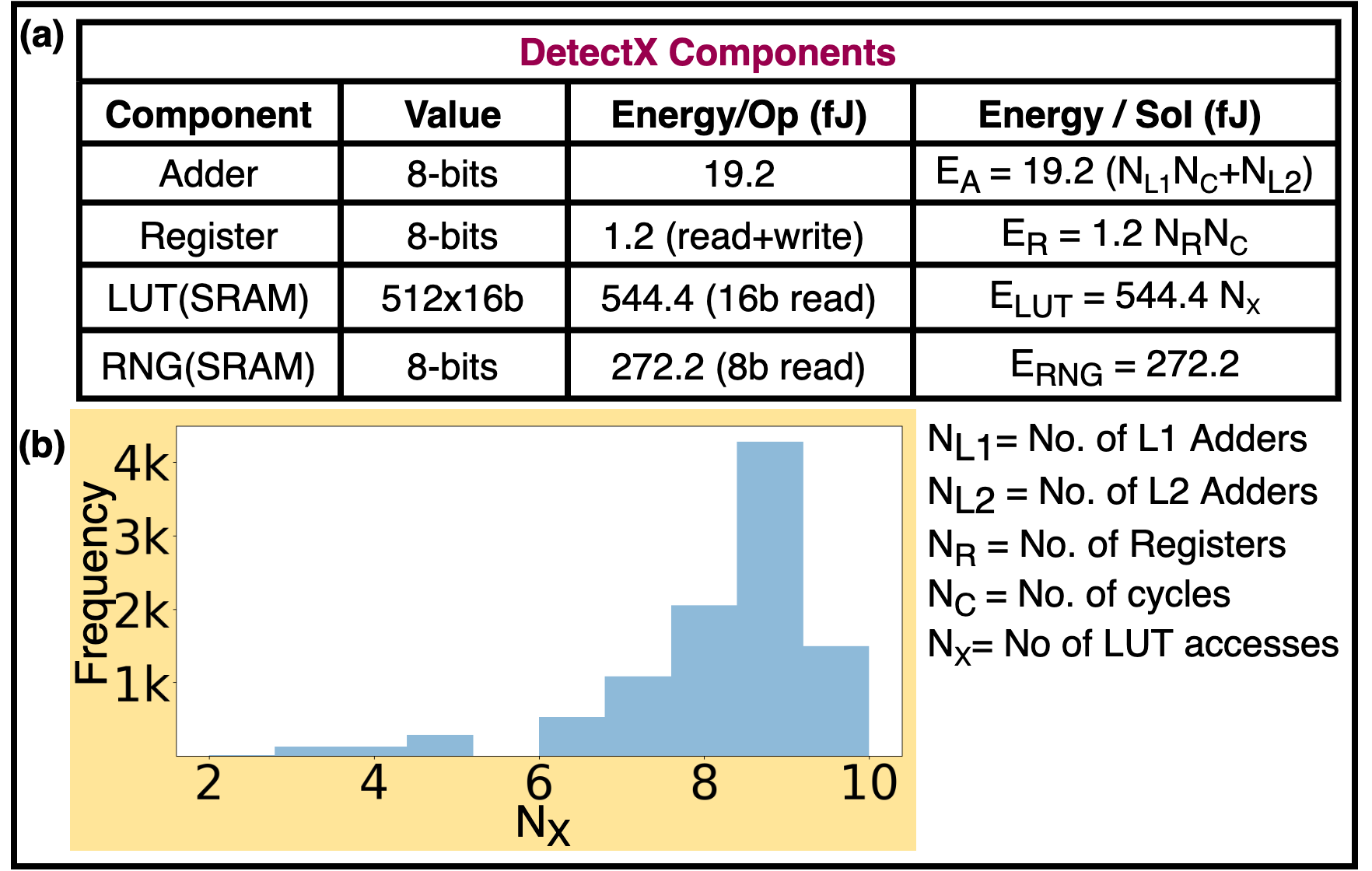}}
    \caption{a) Table showing the Energy/Operation of all the major components in the DetectX module. The Energy/SoI represents the energy required to compute and process a single SoI value. The Energy/SoI value for individual components depend on parameters like $N_{L1}$, $N_{L2}$, $N_R$, $N_C$ and $N_X$. $N_{L1}$, $N_{L2}$, $N_R$, $N_C$ depend on the analog crossbar array parameters shown in Fig. \ref{table_neurosim_impl}a. b) $N_X$ depends on the input SoI value due to the \textit{Binary Search} Operation. The distribution corresponds to the number of LUT accesses for a dataset having equal amounts of clean and adversarial inputs.}
    \label{detectX_energy}
\end{figure}

\subsection{Energy Analysis of the DetectX module }The DetectX module is implemented using 32nm CMOS PTM \cite{balijepalli2007compact}. For energy overhead estimation, we consider the energy consumed by the most significant components only. The energy of all the components of DetectX are evaluated using the Cadence Virtuoso platform. These include the \textit{L1+L2 Adders}, \textit{Registers}, \textit{LUT} and the \textit{RNG} as shown in Fig. \ref{detectX_energy}a. Here,  \textit{Energy/Op} is the energy consumed for a unit operation. For example, the 8-bit \textit{Adder} requires 19.2fJ of energy for one 8-bit addition/subtraction. Similarly, \textit{Energy/SoI} denotes the amount of energy consumed for each input SoI value by the DetectX module. Note, that the \textit{Energy/SoI} value of each component (except the RNG) depends on parameters $N_{L1}, N_{L2}, N_R, N_X$  and $N_{c}$. They denote the number of {\it L1 Adders}, {\it L2 Adders}, {\it Registers}, read cycles and $LUT$ accesses, respectively. The values $N_{L1}, N_{L2}, N_R$ and $N_C$ will be derived in Section \ref{Integration and Hardware Evaluation}. Additionally, for each SoI, the RNG is accessed once. Hence, its \textit{Energy/SoI} is equal to the \textit{Energy/Op}. 

The energies consumed by the \textit{L1+L2 Adders}, \textit{Registers} and the \textit{RNG} are independent of the input SoI values. However, $N_{X}$ depends on the SoI value because of the \textit{Binary Search} operation. Consequently, $E_{LUT}$ depends on the SoI value. To analyse the SoI dependent behaviour of $N_{X}$, we create a dataset of 10k samples having equal number of clean and adversarial inputs. The inputs are fed into the dual-phase trained Neurosim+DetectX model. Corresponding to each input, the SoI values are looked up in the LUT and the number of memory accesses are recorded. The distribution of $N_X$ has been shown in the Fig. \ref{detectX_energy}b.
\begin{figure}
    \centerline{\includegraphics[width=0.5\textwidth]{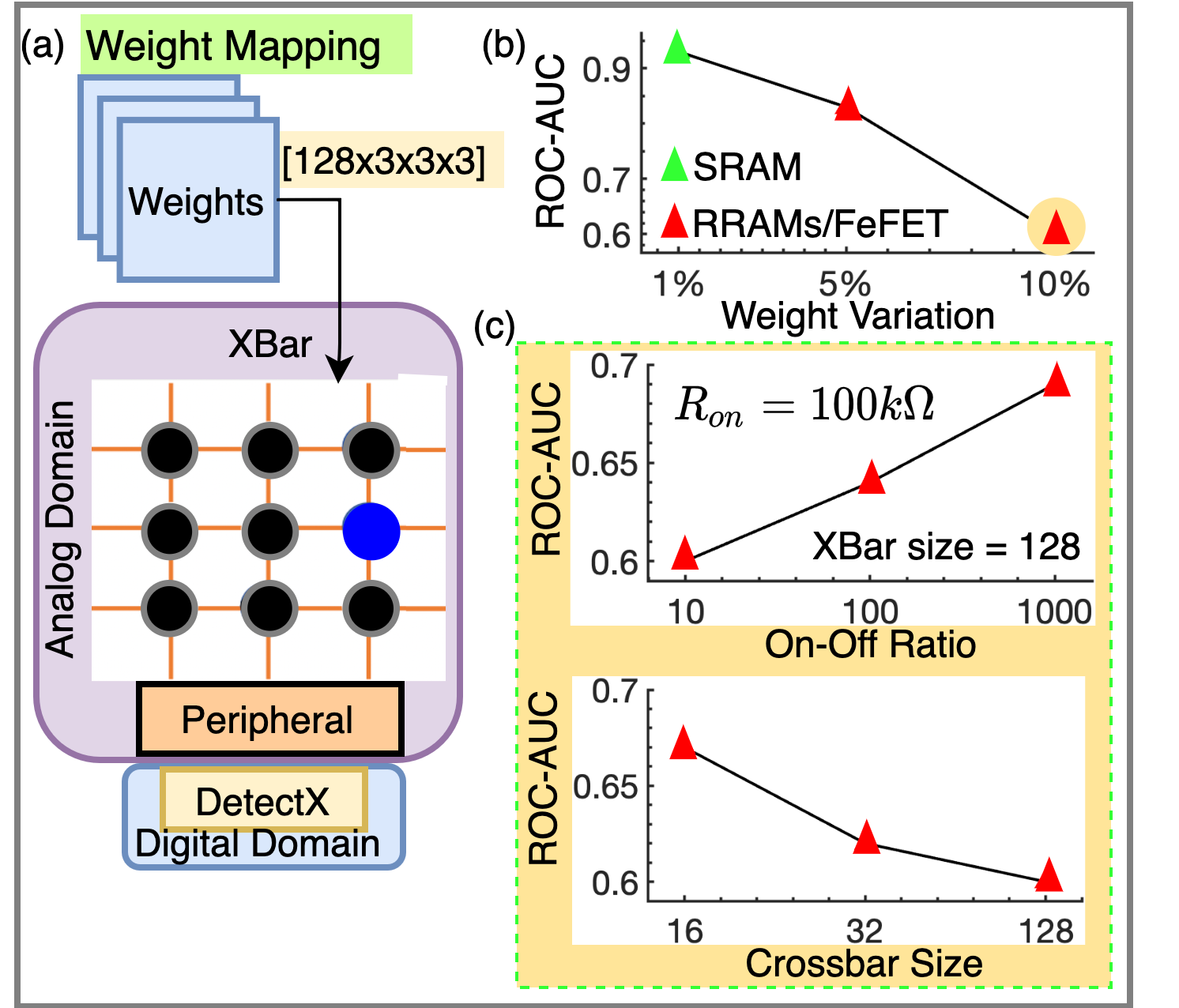}}
    \caption{The overall hardware implementation and evaluation steps of the analog crossbar+DetectX system. a) Shows the mapping of the first layer weights of the VGG8 network (dual-phase trained on CIFAR10 dataset) onto the crossbar array. The crossbar is interfaced with \textit{Peripheral} circuits and the \textit{DetectX} module. b) Shows how the performance of DetectX is affected by the crossbar device non-idealities. The ROC-AUC score is the highest for SRAM based analog crossbar arrays due to extremely small variations in the weights. For NVM devices (RRAMs and FeFETs) the ROC-AUC score is less. c) Analysis of detection performance based on RRAM based analog crossbar architecture. The detection performance is higher for high on-off ratios and small crossbar sizes.}
    \label{neurosim_impl}
\end{figure}
\begin{figure*}
    \centerline{\includegraphics[width=0.75\textwidth]{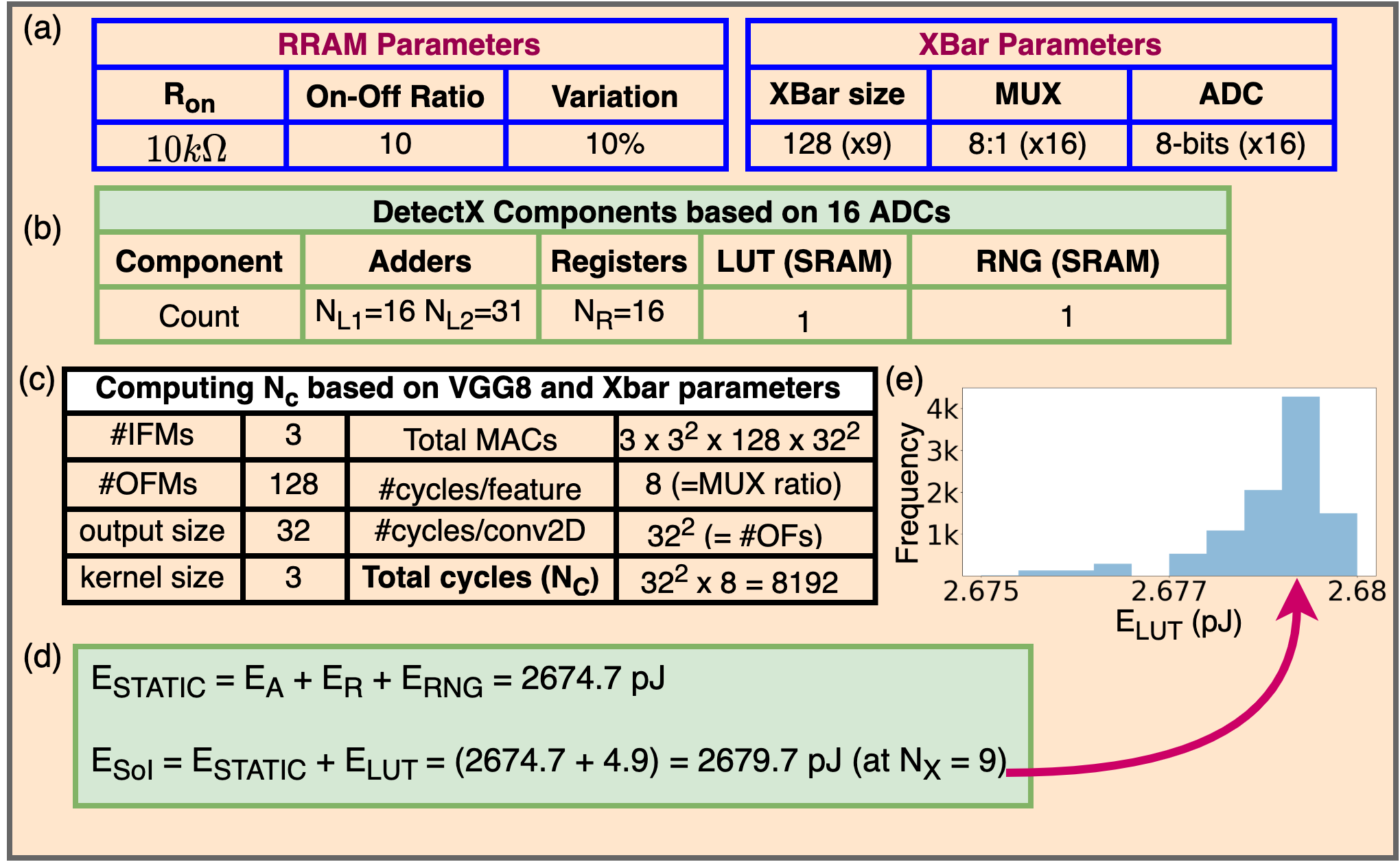}}
    \caption{a) Tables showing the analog crossbar and RRAM device parameters used for hardware evaluation. b) Based on the crossbar configurations (16 output ADCs) as seen in (a), the parameters of the DetectX module are listed. c) The number of cycles required to read all the output features are computed based on the DNN network parameters like number of input and output feature maps (IFMs and OFMs), output size and size of the MUX. d) The data independent $E_{STATIC}$ is computed using $E_A$, $E_R$ and $E_{RNG}$ are computed from the 
    Fig. \ref{detectX_energy}a. The final Energy per SoI $E_{SoI}$ is computed by adding the data independent energy $E_{STATIC}$ and the data dependent LUT access energy $E_{LUT}$. Note, we assume that the number of accesses are 9 while calculating the value of $E_{SoI}$. e) The distribution of the data dependent LUT access energy $E_{LUT}$.}
    \label{table_neurosim_impl}
\end{figure*}
\subsection{Hardware Evaluation of the crossbar+DetectX System}
\label{Integration and Hardware Evaluation}
In this section, we perform hardware evaluations to analyse the performance of the analog crossbar+DetectX system. The hardware evaluation is performed on the Neurosim platform \cite{peng2019dnn+}. Neurosim efficiently maps large-scale DNN architectures (like VGG8, VGG16 \cite{simonyan2014very}) on analog crossbar arrays. Additionally, the platform also considers device conductance variations and non-idealities. This motivates a more realistic hardware evaluation of DNNs. 

Fig. \ref{neurosim_impl} depicts the hardware implementation and evaluation steps used in this section. First, a dual-phase trained VGG8 network (on CIFAR10 dataset) is mapped on a Neurosim-like analog crossbar architecture. The VGG8 network is trained with 8-bit data precision for weights and activations. Fig. \ref{neurosim_impl}a shows the mapping of the first layer weights (dimension 128x3x3x3) onto the analog crossbar array. The analog MAC outputs from the crossbar are converted to digital values by the \textit{Peripheral} circuits. The DetectX module being a digital circuit, is appended at the end of the \textit{Peripheral} stage. 

\textbf{Effects of Crossbar Non-idealities on DetectX's performance: }DetectX is device agnostic, i.e, it can be integrated with different device-based crossbar architectures. However, the detection performance is affected by device non-idealities like variation in weights as shown in Fig. \ref{neurosim_impl}b. Here, for adversarial detection, we use PGD [$\epsilon$=16/255,$\alpha$=2/255,n=10] attacks. For a SRAM device based analog crossbar architecture, the variation of weights is minimal. Hence, we observe a high ROC-AUC score. However, due to significant conductance variation in devices like RRAMs and FeFETs, the detection performance is compromised. 

For all further hardware evaluations we use the RRAM device based analog crossbar architectures. DetectX's performance is sensitive to the crossbar size and the RRAM device on-off ratio as seen in the Fig. \ref{neurosim_impl}c. The on-off ratio of the RRAM device signifies the distinguishability between the high and the low resistance states. Higher on-off ratios lead to better distinction between the clean and adversarial SoI distributions. Hence, a higher ROC-AUC score is observed. Moreover, as the crossbar size is decreased, the overall device non-ideality decreases. This leads to an increase in the ROC-AUC score. Further, DetectX is immune to ADC quantization noise. This is because, DetectX is augmented at the end of the first layer of the DNN. Quantization noise effect is more prevalent in the deeper layers where multiple analog crossbars are used for mapping the DNN layer \cite{chen2018neurosim}.

To improve the performance of DetectX under the impact of hardware non-idealities like device conductance variations, process-voltage-temparature (PVT) variations among others, variation aware training methods \cite{agrawal2019x} can be employed during the dual-phase training. Additionally, new 3D crossbar architectures \cite{eshraghian2021crossstack} and memristive devices \cite{an2020robust} can be utilized to minimize the bit-line parasitic capacitance/resistances and device conductance variations, respectively which can ultimately improve the performance of DetectX.

\textbf{Energy Evaluation of the Crossbar+DetectX system: }For energy evaluation, we use a 128x128 RRAM based crossbar. The parameters of the RRAM device used is shown in Fig. \ref{table_neurosim_impl}a. In Neurosim, each memristive device has 4 analog levels (or 2-bits). Therefore, to implement 8-bit weights, each compute cell in the analog crossbar array contains 4 RRAM devices. Based on the mapping scheme followed by \cite{peng2019dnn+}, each crossbar column outputs 1 output feature map (OFMs). Additionally, in Neurosim, each layer is mapped onto $K^2$ crossbars. Here, $K$ is the kernel dimension. Based on this, the number of crossbars required to map the first layer of the VGG8 network is 9. Each individual crossbar outputs a partial sum value of the total MAC output. The individual partial sums from each crossbar are summed up resulting in 128 final MAC outputs. The \textit{Peripheral} circuit consists of 8:1 MUX and 8-bit ADCs for analog to digital conversion. For the given 128x128 crossbar, we require sixteen 8:1 MUX and sixteen 8-bit ADCs in the \textit{Peripheral} circuit. This has been shown in the Fig. \ref{table_neurosim_impl}a.

The parameters $N_A$,$N_R$,$N_C$ are determined by the number of output ADCs in the \textit{Peripheral} circuit. Based on 16 ADCs, the number of \textit{L1 Adders} $N_{L1}$ is 16. This implies that the number of \textit{L2 Adders} $N_{L2}$ are 16+8+4+2+1 = 31. Likewise, the value of the parameter $N_{R}$ = 16. This has been shown in Fig. \ref{table_neurosim_impl}b. Note that the number of \textit{LUT} and \textit{RNG} do not depend on the \textit{Peripheral} circuit configurations. 

Another factor that determines the energy of the DetectX module is $N_C$. $N_C$ denotes the number of cycles required to sample all the MAC outputs from the analog crossbar. To compute $N_C$, we consider a 32x32x3 sample input from the CIFAR10 dataset. The number of read cycles required to read all the OFMs is equal to the MUX ratio = 8. Additionally, there are 32x32 output features (OFs) corresponding to each OFM. Hence, the effective number of cycles $N_C$ = 32*32*8 = 8192. This has been shown in the Fig. \ref{table_neurosim_impl}c.

The final Energy/SoI value of the DetectX module is calculated using the equation shown in Fig. \ref{table_neurosim_impl}d. The Energy/SoI of the DetectX module consists of two components: $E_{STATIC}$ and $E_{LUT}$. The $E_{STATIC}$ is independent of the input data and is the summation of energies $E_A$, $E_R$ and $E_{RNG}$. Hence, the value of $E_{STATIC}$ is 2674.7 pJ. To compute the energy of the data dependent $E_{LUT}$ value, we assume that $N_X$ = 9. This is the highest number of LUT accesses (for the worst case) as seen in Fig. \ref{table_neurosim_impl}e. Using this, the total Energy/SoI value is 2679.7 pJ.

In the next section, we perform experiments using the Neurosim+DetectX system. Note, that crossbar non-idealities are beyond the scope of this paper. Hence, for all experiments, we assume that the crossbar outputs are free from any kind of device conductance variations. We will perform a detailed analysis of DetectX+non-ideal crossbar system in the future works. 
\section{Experiments and Results}
\subsection{Experiment setup}
We evaluate the crossbar implemented Neurosim+DetectX system using three datasets of varying complexity. First, the CIFAR10 image dataset, which has 60K samples (50k training/10k testing) divided among 10 classes. Second, the CIFAR100 dataset which has 60k samples (50k training/10k testing) divided among 100 classes. Finally, the TinyImagenet dataset having 110k samples (100k training/10k testing) divided among 200 classes. The input dimensions of the samples in CIFAR10/CIFAR100 and TinyImagenet are 32x32 and 64x64, respectively. For the CIFAR10, CIFAR100 and TinyImagenet datasets, we use the VGG8, VGG16 and ResNet18 network architectures, respectively. 

The dual-phase training is performed on the Pytorch framework in an offline manner. For hardware evaluation, we implement the DNN+DetectX system of the Neurosim platform. In all our experiments, we use 8-bit quantization for both DNN weights and activations. For training, we use the Stochastic Gradient Descent (SGD) optimization algorithm. Additionally, for all datasets, we enforce the following hyper-parameters for training [learning rate (LR) = $10^{-3}$ and epochs= 257]. In all the experiments, for Phase1 training, we use the training dataset [clean + PGD ($\epsilon$=16/255, $\alpha$=8/255, $n$= 10)]. While, in the Phase 2 adversarial training, we use training set [clean + PGD ($\epsilon$=4/255, $\alpha$=2/255, $n$=10)]. For creating the SoI-Probability LUT, we use $D_c$ = clean SoIs and $D_a$ = SoI distribution corresponding to PGD [$\epsilon$=8/255, $\alpha$=4/255, $n$=10] inputs.
The adversarial inputs in all the cases are generated offline i,e on the Pytorch framework. The Neurosim framework is used only during the hardware evaluation explained in Section \ref{Integration and Hardware Evaluation}. The value for hyperparameters $\lambda_{c}$ and $\lambda_{a}$ are 0.1 and 0.6, respectively, in all our experiments. 
\subsection{Robustness of the Neurosim+DetectX System}
\label{software evaluation}
\begin{figure}
    \centering
    \includegraphics[width=0.4\textwidth]{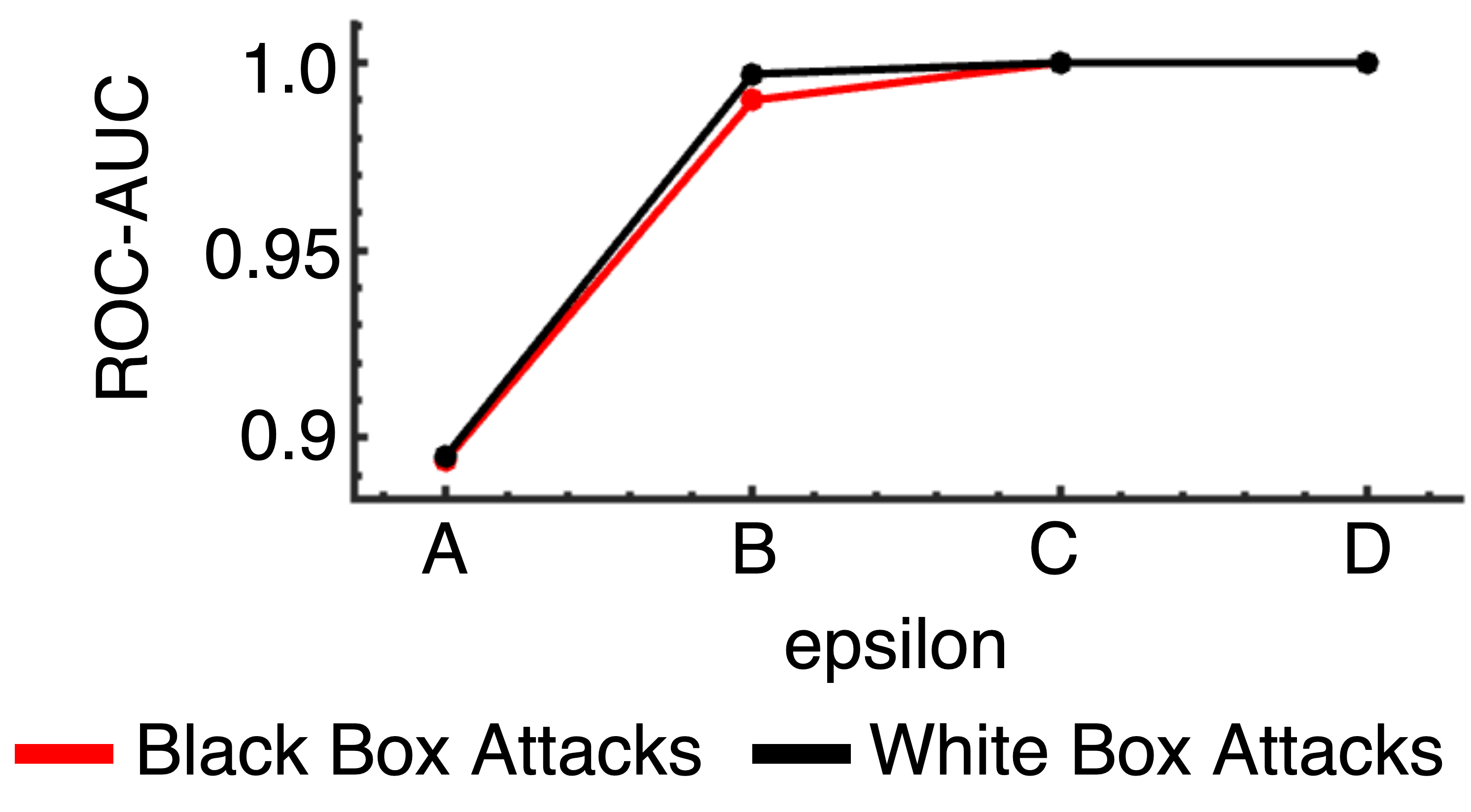}
    
    \caption{DetectX is agnostic towards the type of adversarial attack (WB or BB). Hence, the ROC-AUC curves for both WB and BB attacks follow similar trends. Additionally, for weak adversarial attacks, the ROC-AUC score decreases slightly. Here, A- PGD[8/255,2/255,10], B- PGD[16/255,4/255,10] C- PGD[32/255,4/255,10] D- PGD[32/255,8/255,10]. Although, we show the results obtained for CIFAR10 dataset on a VGG8 network, the trends of the ROC-AUC scores remain constant across different datasets.}
    \label{indep_att}
\end{figure}

In this section, we evaluate the robustness of the Neurosim+DetectX system against different strengths and types of adversarial attacks. For evaluation, we use the metrics \textit{ROC-AUC}, \textit{Error} and \textit{Accuracy} explained in Section \ref{auc_error_accuracy}. Additionally we evaluate the performance of the DetectX module on random gaussian noise attacks. 
\begin{table*}[]
\centering
\caption{Robustness evaluation of the Neurosim+DetectX system with different image datasets and adversarial attacks. We mention the adversarial datasets used for Phase1 and Phase2 training corresponding to each dataset. \textit{D} and \textit{B} denote the Neurosim+DetectX system and baseline model respectively. The baseline model is a DNN trained on clean inputs in the standard manner.}
\begin{tabular}{ccccccc}
\hline
\multicolumn{7}{c}{\textbf{CIFAR10 VGG8 Phase1-[$\epsilon$= 32/255, $\alpha$= 8/255, n=10], Phase2- [$\epsilon$= 4/255, $\alpha$= 2/255, n= 10]}}                                                                                                                                                                                                                                                                                                                        \\ \hline
\multicolumn{1}{c|}{\multirow{2}{*}{\textbf{Attack}}} & \multicolumn{2}{c|}{\textbf{FGSM}}                                             & \multicolumn{4}{c}{\textbf{PGD}}                                                                                                                                                              \\
\cline{2-7} 
\multicolumn{1}{c|}{}                                 & \multicolumn{1}{c|}{\textbf{e: 16/255}} & \multicolumn{1}{c|}{\textbf{e: 32/255}} & \multicolumn{1}{c|}{\textbf{8/255, 2/255, 10}} & \multicolumn{1}{c|}{\textbf{8/255, 4/255, 10}} & \multicolumn{1}{c|}{\textbf{16/255, 4/255, 10}} & \multicolumn{1}{c}{\textbf{32/255, 8/255, 10}}                          \\ \cline{1-7} 
\multicolumn{1}{c|}{\textbf{ROC-AUC}}                 & \multicolumn{1}{c|}{0.97}             & \multicolumn{1}{c|}{0.99}              & \multicolumn{1}{c|}{0.895}                    & \multicolumn{1}{c|}{0.99}                    & \multicolumn{1}{c|}{1}                      & \multicolumn{1}{c}{1}                       
\\ \hline
\multicolumn{1}{c|}{\textbf{Accuracy D / B}}           & \multicolumn{6}{c}{80.1 / 87}                                                                                                                                                                                                                                                                               \\ \hline
\multicolumn{1}{c|}{\textbf{WB Error D / B}}           & \multicolumn{1}{c|}{1.7 / 35.3}                 & \multicolumn{1}{c|}{1.1 / 48.3}                  & \multicolumn{1}{c|}{9.1 / 53.2}                 & \multicolumn{1}{c|}{1.5 / 61.9}               & \multicolumn{1}{c|}{0 / 66.17}               & \multicolumn{1}{c}{0 / 67.18}                                      \\ \hline
\multicolumn{1}{c|}{\textbf{BB Error D / B}}           & \multicolumn{1}{c|}{1.6 / 63.8}                 & \multicolumn{1}{c|}{1.4 / 72.6}                  & \multicolumn{1}{c|}{8.2 / 36.1}             & \multicolumn{1}{c|}{1.2 / 66.32}                & \multicolumn{1}{c|}{0 / 72.31}                  & \multicolumn{1}{c}{0 / 79.39}                                       \\ \hline
\multicolumn{7}{c}{\textbf{CIFAR100 VGG16 Phase1-[$\epsilon$= 32/255, $\alpha$= 8/255, n=10], Phase2- [$\epsilon$= 4/255, $\alpha$= 2/255, n= 10]}}                                                                                                                                                                                                                                                                                                                      \\ \hline
\multicolumn{1}{c|}{\textbf{ROC-AUC}}                 & \multicolumn{1}{c|}{0.98}             & \multicolumn{1}{c|}{0.99}              & \multicolumn{1}{c|}{0.99}                    & \multicolumn{1}{c|}{1}                     & \multicolumn{1}{c|}{1}                      & \multicolumn{1}{c}{1}                     
\\ \hline
\multicolumn{1}{c|}{\textbf{Accuracy D / B}}           & \multicolumn{6}{c}{50.1 / 60.2}                                                                                                                                                                                                                                                                                \\ \hline
\multicolumn{1}{c|}{\textbf{WB Error D / B}}           & \multicolumn{1}{c|}{5.1 / 59.4}                 & \multicolumn{1}{c|}{2.16 / 67.24}                  & \multicolumn{1}{c|}{6.7 / 94.13}             & \multicolumn{1}{c|}{0 / 96.66}              & \multicolumn{1}{c|}{0 / 97.76}                 & \multicolumn{1}{c}{0 / 97.77}                                         \\ \hline
\multicolumn{1}{c|}{\textbf{BB Error D / B}}           & \multicolumn{1}{c|}{4.9 / 55.6}                 & \multicolumn{1}{c|}{1.23 / 68.9}                  & \multicolumn{1}{c|}{5.3 / 92.46}              & \multicolumn{1}{c|}{0 / 95.3}             & \multicolumn{1}{c|}{0 / 96.9}              & \multicolumn{1}{c}{0 / 98.2}                                      \\ \hline
\multicolumn{7}{c}{\textbf{TinyImagenet ResNet18 Phase1-[$\epsilon$= 32/255, $\alpha$= 8/255, n=10], Phase2- [$\epsilon$= 4/255, $\alpha$= 2/255, n= 10]}}                                                                                                                                                                                                                                                                                                               \\ \hline
\multicolumn{1}{c|}{\textbf{ROC-AUC}}                 & \multicolumn{1}{c|}{0.84}             & \multicolumn{1}{c|}{0.92}              & \multicolumn{1}{c|}{0.65}                    & \multicolumn{1}{c|}{0.86}                    & \multicolumn{1}{c|}{0.98}                     & \multicolumn{1}{c}{1}                                         \\
 \hline
\multicolumn{1}{c|}{\textbf{Accuracy D / B}}           & \multicolumn{6}{c}{42.2 / 53.5}                                                                                                                                                                                                                                                                           \\ \hline
\multicolumn{1}{c|}{\textbf{WB Error D / B}}           & \multicolumn{1}{c|}{13.2 / 56.5}      & \multicolumn{1}{c|}{6.1 / 63.21}                  & \multicolumn{1}{c|}{59.4 / 88}              & \multicolumn{1}{c|}{12.36 / 90.7}             & \multicolumn{1}{c|}{10.1 / 91.2}              & \multicolumn{1}{c}{0.1 / 93}                                       \\ \hline
\multicolumn{1}{c|}{\textbf{BB Error D / B}}           & \multicolumn{1}{c|}{10.17 / 55.47}                 & \multicolumn{1}{c|}{5.2 / 60.27}                  & \multicolumn{1}{c|}{53.6 / 76.3}                        & \multicolumn{1}{c|}{11.1 / 90.2}                         & \multicolumn{1}{c|}{5.4 / 90.7}                          & \multicolumn{1}{c}{1 / 91.9}                                                  \\ \hline
\end{tabular}
\label{table}
\end{table*}

In Fig. \ref{indep_att}, we detect WB and BB PGD attacks of varying strengths using the Neurosim+DetectX system trained on CIFAR10 dataset and VGG8 architecture. Here, the strength of the adversarial attacks A, B, C and D increase from left to right. We find that the ROC-AUC curves for WB and BB attacks have similar trends. This shows that \textbf{DetectX is agnostic towards WB and BB adversarial attacks.} 


\begin{figure}
\centerline{\includegraphics[width=0.4\textwidth]{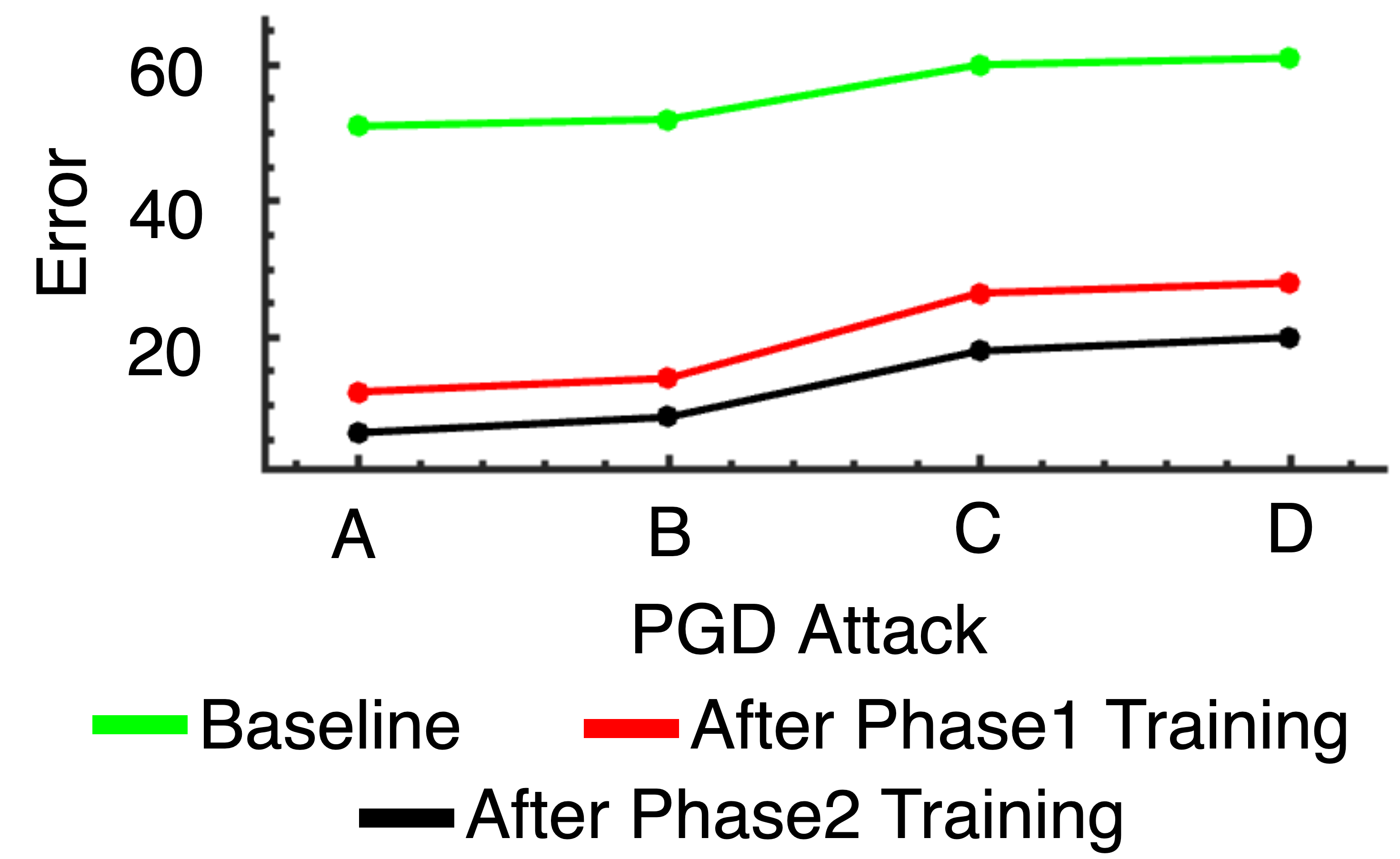}}
\caption{Performance of DetectX with weak PGD attacks. Here, we use Neurosim+DetectX system trained on the CIFAR10 dataset and VGG8 architecture. A-PGD[3/255,0.5/255,10], B-PGD[4/255,0.5/255,10], C-PGD[3/255,1/255,10] and D-PGD[4/255,1/255,10]. After Phase1 training, the \textit{Error} is significantly lowered than the \textit{Baseline} model. The Phase2 training further lowers the \textit{Error} value due to adversarial training.}
\label{weak_attacks}
\end{figure}

Table \ref{table} shows the overall robustness of the Neurosim+DetectX system for different datasets. For each dataset, we mention the DNN architecture, Phase1 and Phase2 adversarial data on the top. The notations \textit{D} and \textit{B} stand for Neurosim+DetectX system and the baseline model, respectively. The baseline model is a DNN trained in the standard manner on clean data. We report a single ROC-AUC for both WB and BB attacks because of the attack agnostic property of DetectX. Additionally, the \textit{Accuracy} is not affected by the type and strength of adversarial attacks. This is because $D_c$ and the \textit{SoI-Probability LUT} remains constant for each dataset. However, \textit{Error} strongly depends on the type and strength of the attack. Hence, we individually report the \textit{Error} values corresponding to WB and BB attacks.

For generating WB attacks, we use the cross-entropy loss function (Equation \ref{ce_loss}) of the Neurosim+DetectX system. For generating BB attacks, we use the following dataset and DNN combinations: (CIFAR10, VGG16), (CIFAR100, ResNet18) and (TinyImagenet, VGG16). All the networks are trained in the standard manner on clean data.  

Overall, for both FGSM and PGD attacks, the Neurosim+DetectX system has a high ROC-AUC score. Consequently, this leads to very low \textit{Error} values for both WB and BB attacks when compared to the baseline model. Additionally, the performance with CIFAR10 and CIFAR100 datasets is better than the TinyImagenet dataset. This is because of the wide data distribution of the TinyImagenet dataset. For weaker adversarial attacks, the ROC-AUC score is slightly lower than the stronger attacks. This increases the \textit{Error} value. However, with Phase2 adversarial training the increase in the \textit{Error} value is compensated. Additionally, the dual-phase training also causes a drop in \textit{Accuracy} compared to the baseline model. 

Fig. \ref{weak_attacks} demonstrates the efficacy of the dual-phase training. Here, weak WB PGD attacks are launched on the Neurosim+DetectX system trained on the CIFAR10 dataset and VGG8 architecture. It can be seen that after Phase1 training, the \textit{Error} value drops significantly compared to the baseline model. With Phase2 adversarial training, the \textit{Error} is further lowered. Thus, despite a slight reduction in the detection performance for weak attacks, the \textit{Error} values remain considerably low.

\begin{figure}
\centerline{\includegraphics[width=0.4\textwidth]{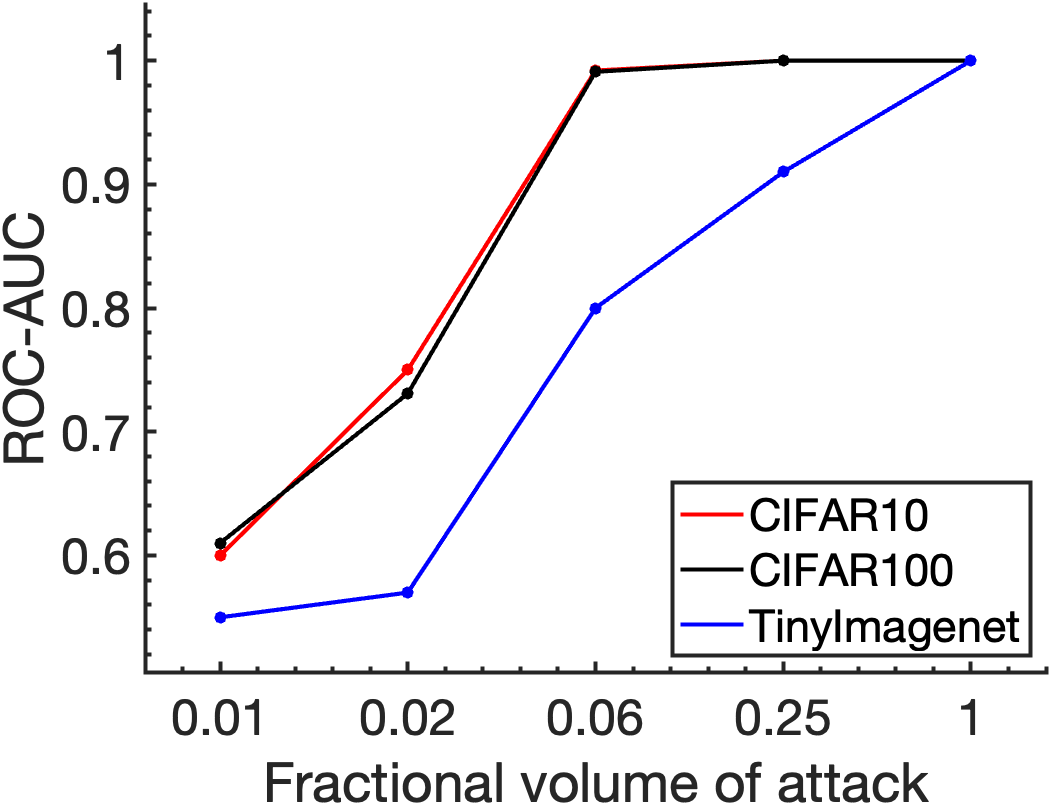}}
\caption{ROC-AUC scores corresponding to different fractional volumes of random gaussian noise attacks. The starting positions for the patch are randomly selected. For larger fractional volumes of attack the detector has higher ROC-AUC score.}
\label{random}
\end{figure}

\textbf{Performance with random noise attacks: }DetectX is not only efficient in detecting gradient based attacks, but also achieves significant performance for non-gradient based attacks. We demonstrate this by adding random gaussian noise patch to the input. We define patch size as the fractional volume of the input 3D tensor. Fig. \ref{random} shows the ROC-AUC scores of the detector corresponding to different fractional volumes of attack for CIFAR10, CIFAR100 and TinyImagenet datasets. Not only does the detector detect random gaussian noise attacks on the full volume, but it can also effectively detect attacks with a high ROC-AUC score when the fractional volume is reduced. Further, due to higher dataset complexity, TinyImagenet has an overall lower ROC-AUC score for smaller fractional volumes. 



\subsection{Comparison with Previous Works}
\begin{figure*}
\centerline{\includegraphics[width=1\textwidth]{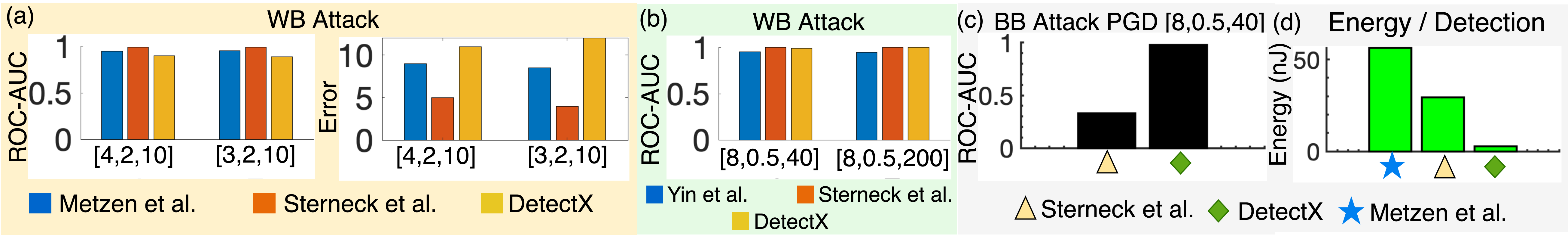}}
\caption{Comparison of detection performance and energy consumption of the Neurosim+DetectX system with other state-of-the-art works. Note, PGD[x,y,z] means $\epsilon$=x/255, $\alpha$=y/255 and n= z. a). Comparison of ROC-AUC scores and Error values for CIFAR10 dataset implemented on the VGG16 Network. It can be seen that with weak WB attacks, DetectX has a slightly lower performance when compared to Metzen et al. \cite{metzen2017detecting} and Sterneck et al. \cite{sterneck2021noise}. However, due to Phase2 training, the \textit{Error} values are maintained at low values. b) With strong WB attacks, DetectX outperforms both Sterneck et al. \cite{sterneck2021noise} and Yin et al. \cite{yin2019gat}. c) Due to attack agnostic behaviour of DetectX, the performance is equally high for BB attacks. d) Finally, DetectX consumes more than 10x lower energy when compared to other state-of-the-art works. The neural network based detectors of Metzen et al. \cite{metzen2017detecting} and Sterneck et al. \cite{sterneck2021noise} are implemented on the Neurosim platform for energy evaluation. The energy of the DetectX module is computed using the procedure discussed in Section \ref{Integration and Hardware Evaluation}.}
\label{energy_comparison}
\end{figure*}
  
In this section, we compare the robustness and energy efficiency the Neurosim+DetectX system with previous state-of-the-art works on adversarial input detection \cite{metzen2017detecting,sterneck2021noise, yin2019gat}. All these works employ neural network based detectors to achieve state-of-the-art performance. 

From the perspective of adversarial robustness, we examine the consequences of two interesting features of DetectX: 1) Indifference towards WB and BB adversarial attacks, 2) Immunity towards dynamic adversarial attacks. In Fig. \ref{energy_comparison}a, we perform dual-phase training of a VGG16 network on the CIFAR10 dataset. Then we compare the detection performance of the Neurosim+DetectX system with different works under weak PGD attacks. These include PGD[$\epsilon$= 4/255, $\alpha$=2/255, n=10] and PGD[$\epsilon$= 3/255, $\alpha$=2/255, n=10]. The ROC-AUC score of the Neurosim+DetectX system is slightly lower than Metzen et al. \cite{metzen2017detecting} and Sterneck et al. \cite{sterneck2021noise}. However, due to Phase2 adversarial training, we see that the overall \textit{Error} value is sufficiently lowered and comparable to the previous works. Next, in Fig. \ref{energy_comparison}b, we find that for strong PGD attacks, PGD[$\epsilon$= 8/255, $\alpha$=0.5/255, n=40] and PGD[$\epsilon$= 8/255, $\alpha$=0.5/255, n=200]) DetectX performs significantly better than the works by Yin et al. \cite{yin2019gat} and Sterneck et al. \cite{sterneck2021noise} with a ROC-AUC score of 1. Due to the high detection performance, we do not show the \textit{Error} values here. This is because all the adversarial inputs are detected and \textit{Error} will be zero. In Fig. \ref{energy_comparison}c, we compare DetectX's performance under BB (PGD [$\epsilon$= 8/255, $\alpha$=0.5/255, n=40]) attack with Sterneck et al. \cite{sterneck2021noise} DetectX has a ROC-AUC score of 0.99 while Sterneck et al. \cite{sterneck2021noise} shows a ROC-AUC score of 0.37. A ROC-AUC score of less than 0.5 suggests that the classification results are flipped and the detection is not reliable. This demonstrates the attack agnostic feature of DetectX which implies increased robustness. Note, other adversarial detection works do not report BB attacks. Finally, Metzen et al. \cite{metzen2017detecting} and Sterneck et al. \cite{sterneck2021noise} show that neural network based detectors are prone to \textit{dynamic adversarial attacks} which can lead to a degradation in the detection performance. DetectX being a non-neural network based detector is immune to dynamic adversarial attacks. This makes the Neurosim+DetectX system more robust compared to other DNN+neural network-based systems. 

\textbf{Energy Comparison with Previous Works: }Having discussed the adversarial robustness aspects, we now compare the energy consumption of the DetectX module when compared to previous works. For this, the neural network based detectors of Metzen et al. \cite{metzen2017detecting} and Sterneck et al. \cite{sterneck2021noise} are implemented on the Neurosim platform followed by energy evaluation. Clearly, as seen in Fig. \ref{energy_comparison}c, DetectX requires 10x to 25x less energy for adversarial input detection when compared to previous works.

Essentially, DetectX is one of the first works that uses hardware signatures like the Sum of column Currents (SoI) in analog crossbar arrays for adversarial detection. Since DetectX does not require any neural network based-detector, it is highly energy efficient compared to the previous state-of-the-art works. At the same time, DetectX is also proficient in defending the DNN against various gradient and non-gradient based adversarial attacks of different strengths. Finally, the DetectX approach is device agnostic. Meaning, that DetectX can be augmented with analog crossbar arrays irrespective of the memristive device used.

\section{Conclusion}
In this work, we propose a hardware-centric adversarial detection strategy, DetectX. We show that hardware-based signatures like Sum of column Currents (SoIs) in analog memristive crossbar arrays can be used to detect adversarial inputs. For this, we use a dual-phase training approach to increase the separation between clean and adversarial SoIs. We implement the DetectX module using 32nm CMOS Predictive Technology Model. The DetectX module consists of a SoI computing unit and a simple Look-Up Table based detector. We integrate the DetectX module with the Neurosim crossbar evaluation platform. Our experiments on benchmark datasets show that DetectX is highly energy efficient and achieves state-of-the-art detection performance against strong white box and black box attacks in comparison to previous works. Thus, DetectX is an energy efficient and robust solution for adversarial input detection on hardware. 

\section{Acknowledgement}
This work was supported in part by C-BRIC, Center for Brain-inspired Computing, a JUMP center sponsored by DARPA and SRC, the National Science Foundation (Grant\#1947826), the Technology Innovation Institute, Abu Dhabi and the Amazon Research Award.

\bibliographystyle{IEEEtran}
\bibliography{DetectX_final.bib}

\end{document}